\documentclass[twocolumn, amsmath, amssymb]{revtex4-2}

\makeatletter
\def\makeLineNumberRight{%
  \ifNAT@numbersleft
    \def\@LineNumber{\hb@xt@\linenumberwidth{\hss\arabic{linenumber}}}%
    \reversemarginpar
  \else
    \def\@LineNumber{\hb@xt@\linenumberwidth{\arabic{linenumber}\hss}}%
    \normalmarginpar
  \fi
  \ifnum\value{linenumber}>0
    \marginpar{\hskip\columnwidth\hskip-\marginparsep\@LineNumber}%
  \fi
}
\makeatother

\usepackage{graphicx}
\usepackage{dcolumn}
\usepackage{bm}
\usepackage[mathlines]{lineno}

\begin{document}

\preprint{APS/123-QED}

\title{DomiRank Centrality: revealing structural fragility of \\ complex networks via node dominance}

\author{Marcus Engsig}
 \email{marcus.w.engsig@gmail.com}

\affiliation{%
Directed Energy Research Centre, Technology Innovation Institute, Abu Dhabi, United Arab Emirates.
}%

\author{Alejandro Tejedor}%
 \email{alej.tejedor@gmail.com}
\affiliation{%
 Institute for Biocomputation and Physics of Complex Systems (BIFI), Universidad de Zaragoza, 50018 Zaragoza, Spain
}%
\affiliation{
 Department of Theoretical Physics, University of Zaragoza, Zaragoza 50009, Spain
}%

\affiliation{%
 Department of Civil and Environmental Engineering, University of California Irvine, Irvine, CA 92697, USA.
}%

\author{Yamir Moreno}
\email{yamir.moreno@gmail.com}
\affiliation{%
 Institute for Biocomputation and Physics of Complex Systems (BIFI), University of Zaragoza, 50018 Zaragoza, Spain
}%
\affiliation{
 Department of Theoretical Physics, University of Zaragoza, Zaragoza 50009, Spain
}%
\affiliation{CENTAI Institute, Turin 10138, Italy.
}%

\author{Efi Foufoula-Georgiou}
\email{efi@uci.edu}
\affiliation{%
 Department of Civil and Environmental Engineering, University of California Irvine, Irvine, CA 92697, USA.
}%
\affiliation{
 Department of Earth System  Science, University of California Irvine, Irvine, CA 92697, USA.
}%

\author{Chaouki Kasmi}
\email{chaouki.kasmi@tii.ae}
\affiliation{%
Directed Energy Research Centre, Technology Innovation Institute, Abu Dhabi, United Arab Emirates
}%

\date{\today}

\begin{abstract}
Determining the key elements of interconnected infrastructure and complex systems is paramount to ensure system functionality and integrity. This work quantifies the dominance of the networks' nodes in their respective neighborhoods, introducing a novel centrality metric, DomiRank, that integrates local and global topological information via a tunable parameter. We present an analytical formula and an efficient parallelizable algorithm for DomiRank centrality, making it applicable to massive networks. From the networks' structure and function perspective, nodes with high values of DomiRank highlight fragile neighborhoods whose integrity and functionality are highly dependent on those dominant nodes.  Underscoring this relation between dominance and fragility, we show that DomiRank systematically outperforms other centrality metrics in generating targeted attacks that effectively compromise network structure and disrupt its functionality for synthetic and real-world topologies. Moreover, we show that DomiRank-based attacks inflict more enduring damage in the network, hindering its ability to rebound and, thus, impairing system resilience. DomiRank centrality capitalizes on the competition mechanism embedded in its definition to expose the fragility of networks, paving the way to design strategies to mitigate vulnerability and enhance the resilience of critical infrastructures.
\end{abstract}

\maketitle

\section{Introduction}

Complex systems consist of many interacting components, with dynamics and emergent behavior being system properties.  However, not all the constituents of such systems are equivalently central to their structure and dynamics, and in some systems, a few elements might be critical to ensure the integrity of the complex systems' structure or functionality \cite{Albert2000,Callaway2000,jeong2001lethality,de2014navigability,boccaletti2006complex, Goltsev2012,Gao2016, CruaAsensio2017,Farooq2019,Guilbeault2021}. Our capacity to accurately and efficiently identify key elements of such complex systems is at the core of actions as diverse as providing the most suitable website on an internet search \cite{page1999pagerank}, defining a vaccination scheme to mitigate the spreading of a disease \cite{Kitsak2010,Salathe2010,Wang2016,Pung2022}, or ensuring the integrity and functionality of transportation networks and critical infrastructures \cite{Guimera2005, Wu2006, brown2006defending, Carvalho2009, Duan2014}. 

Network theory, by abstracting complex systems as a collection of nodes (system constituents) and links (interactions), has been instrumental in providing a general framework to assess different aspects of the relative importance of nodes in a network,  yielding different node centrality definitions depending on the evaluated aspects, ranging from considering only the number of links a node has (degree centrality), aggregating the importance of a node's neighborhood (e.g., eigenvector \cite{bonacich1972factoring}, Katz \cite{katz1953new}, and PageRank \cite{page1999pagerank} centralities) to considering the relative position of the node in the network (e.g., closeness and betweenness \cite{freeman1977set} centralities) or the role of the node in a dynamic process (e.g., current-flow \cite{brandes2005centrality}, entanglement\cite{ghavasieh2021unraveling}, and random-walk \cite{PhysRevE.90.032812} centralities). The performance of these centralities is often benchmarked against each other in evaluating their capacity to generate targeted attacks to dismantle the network's structure or disrupt its functionality. In fact, centrality metrics have a pivotal role in designing mitigation strategies to enhance network robustness and resilience, critical emerging properties of utmost importance to maintain our day-to-day privileges and necessities, which heavily rely on interconnected infrastructures such as the internet \cite{Albert2000, Sole2008, doyle2005robust} or the power grid \cite{rinaldi2001identifying, Cohen20001, Schafer2018}.
 
 In this work, we propose a novel centrality, the `DomiRank' centrality. Intuitively, it quantifies the degree of dominance of nodes in their respective neighborhoods. Thus, high scores of DomiRank centrality are associated with nodes surrounded by a large number of unimportant (e.g., typically low-degree) nodes, which they dominate.  This new centrality gives importance to nodes based on how locally {\it dominant} they are, where the extent of the \textit{dominance} effect can be modulated through a tuneable parameter ($\sigma$). Contrary to other centralities such as eigenvector or PageRank, and due to an implicit competition mechanism in the definition of DomiRank, connected nodes tend to have more disparate scores in terms of DomiRank centrality. We demonstrate that the inherent properties of DomiRank make both synthetic and real-world networks particularly fragile to the DomiRank centrality-based attacks,  outperforming all other centrality-based attacks. Furthermore, we show that the DomiRank-based attack consistently outperforms most of the computationally feasible iterative (recomputed after each node removal) attack methods (e.g., degree, PageRank), and it causes more enduring damage than the efficient iterative betweenness attack.  We provide both an analytical formula and a computationally efficient iterative algorithm for DomiRank, enabling it to be computed on graphical processing units (GPUs) with a parallelizable computational cost scaling with the number of links, allowing the centrality to be computed for massive sparse networks.

\section{DomiRank Centrality}

\subsection{Definition and interpretation}
We define DomiRank centrality, denoted $\boldsymbol{\Gamma}\in \mathbb{R}_{N\times 1}$, as the stationary solution of the following dynamical process  

\begin{equation}
\frac{d \boldsymbol{\Gamma}(t)}{d t} = \alpha A(\theta\boldsymbol{1}_{N\times 1} - \boldsymbol{\Gamma}(t)) - \beta \boldsymbol{\Gamma}(t),
\label{eq:intuition}
\end{equation}

\noindent where $A\in\mathbb{R}_{N\times N}$ is the adjacency matrix of the network $\mathcal{N}$ and $\{\alpha, \beta, \theta \in \mathbb{R} : \lim_{t\to\infty} \boldsymbol{\Gamma}(t) = \boldsymbol{\Gamma}\in\mathbb{R}_{N\times 1}\}$. Note that the definition presented here is valid for unweighted, weighted, directed, and undirected networks, so in the more general case, a non-zero entry of the adjacency matrix $A_{ij}=w_{ij}$ represents the existence of a link from node $i$ to node $j$ with a weight $w_{ij}$.
By expanding the term $\alpha A(\theta\boldsymbol{1}_{N\times 1} - \boldsymbol{\Gamma}(t))$, we obtain that the rate of change $\frac{d \boldsymbol{\Gamma}(t)}{d t}$ has a positive contributing term proportional to the nodal degree $\boldsymbol{k}=A\boldsymbol{1}_{N\times 1}$, and two negative contributing terms: the first proportional to the sum of $\boldsymbol{\Gamma}(t)$ over each node's neighbors ($ A\boldsymbol{\Gamma}(t)$), and the second proportional to the current value of $ \boldsymbol{\Gamma}(t)$.  Thus, for the $i-$th node, Eq. \ref{eq:intuition} reads  

\begin{equation}
\frac{d \Gamma_i(t)}{d t} = \alpha (\theta k_i - \sum_j w_{ij} \Gamma_j(t)) - \beta \Gamma_i(t). 
\label{eq:iNode_general}
\end{equation}

For enhanced interpretability and without loss of generality, we discuss the case of an unweighted network, for which Eq \ref{eq:iNode_general} reduces to

\begin{equation}
\frac{d \Gamma_i(t)}{d t} =  \alpha \sum_{j \in \textrm{neighbors}_i} \Big [ \theta-\Gamma_j (t)) \Big ] - \beta \Gamma_i(t).
\label{eq:iNode_uw}
\end{equation}

\noindent where neighbors$_i$ refers to the set of nodes directly connected to node $i$.

From a simple model perspective, $\boldsymbol{\Gamma}(t)$ can be interpreted as the evolving fitness of the individuals in a population subject to competition. Two different processes can alter the fitness of each individual: $(i)$ {\it Natural relaxation} - fitness naturally converges to zero at a rate proportional to a constant $\beta$; $(ii)$   {\it  Competition} – Individuals compete with each neighbor for a limited amount of resources, with their fitness reflecting their capacity to successfully maintain those resources. An individual's fitness tends to increase by being connected to neighbors whose fitness are below the threshold for domination ($\theta$) and decreases otherwise. Thus, the fitness of each individual changes proportionally to $(\sum_{j \in \textrm{neighbors}} \theta-\Gamma_j (t))$, where the proportionality constant is denoted by $\alpha$ and represents the degree of competition between neighboring individuals. 

Notably, the fitness score of a given individual $k$ is a function of $(i)$ {\it its number of neighbors:} the larger the number of neighbors of $k$, the more resources at stake, and therefore the larger the potential of $k$ to increase/decrease its fitness, and; $(ii)$ {\it its neighbors’ neighborhood:} having neighbors lacking dominance in their respective neighborhoods due to either the absence of neighbors or the presence of dominant neighbors increases the fitness of individual $k$. In other words, a given individual results in having a high value of fitness via the dominance of its neighborhood, either due to the direct dominance of its neighbors (quasi-solitary individuals) or via collusion (joint dominance) emerging from the synergetic action of several individuals in suppressing the fitness of a common neighbor while incrementing their respective fitness.
The DomiRank centrality is thus based on the concept of dominance to provide scores to nodes that contextualize their importance in their neighborhood. Consequently, its direct interpretation in systems wherein interactions are mediated by dominance/power-based relations, such as Rich-Club networks \cite{colizza2006detecting, mcauley2007rich, van2011rich, Alstott2014}, is apparent. In the Supporting Material (SM - see Section S-I), we provide illustrative examples of Rich-Club networks that shed more light on the relevance of the key factors controlling the emergence of dominant and dominated nodes via the joint dominance (collusion) mechanism and how to steer the relative power exploiting the concept of joint dominance.

From Eq. \ref{eq:intuition}, we note that the centrality converges when $\alpha A(\theta\boldsymbol{1}_{N\times 1} - \boldsymbol{\Gamma}(t)) = \beta \boldsymbol{\Gamma}(t)$, for which the analytical expression (see appendix for proof) of the DomiRank centrality $\boldsymbol{\Gamma}\in \mathbb{R}_{N\times 1}$ is given by: 

\begin{equation}
\boldsymbol{\Gamma} = \theta \sigma (\sigma A + I_{N\times N})^{-1}  A  \boldsymbol{1}_{N\times 1},
\label{eq:analytic}
\end{equation}
where $\{ \sigma = \frac{\alpha}{\beta} \in \mathbb{R} : \det(\sigma A + I_{N\times N}) \neq 0\}$. A convergence interval can be defined for $\sigma$, such that it is bounded  as follows:
\begin{equation}
\sigma(\mathcal{N}) \in \left(0, \frac{1}{-\lambda_N}\right),
\label{eq:supremum}
\end{equation}
where $\lambda_N$ represents the minimum (dominant negative) eigenvalue of $A$. Also note that the threshold for domination, $\theta$, only plays a rescaling role on the resulting DomiRank centrality, and therefore, we choose $\theta=1$ without loss of generality.

We recall that the definition, interpretation, and use of DomiRank are valid for both undirected and directed networks. Only note that in the case of directed graphs, the adjacency matrix used in the definition of DomiRank (e.g., Eq. \ref{eq:analytic}) should correspond to the reverse of the graph relevant for the transfer of resources (e.g., information, traffic, etc.) to be consistent with the underlying concept of dominance.

\subsection{The role of DomiRank's parameter $\sigma$}

\begin{figure}[t]
    \centering
    \includegraphics[width = \linewidth]{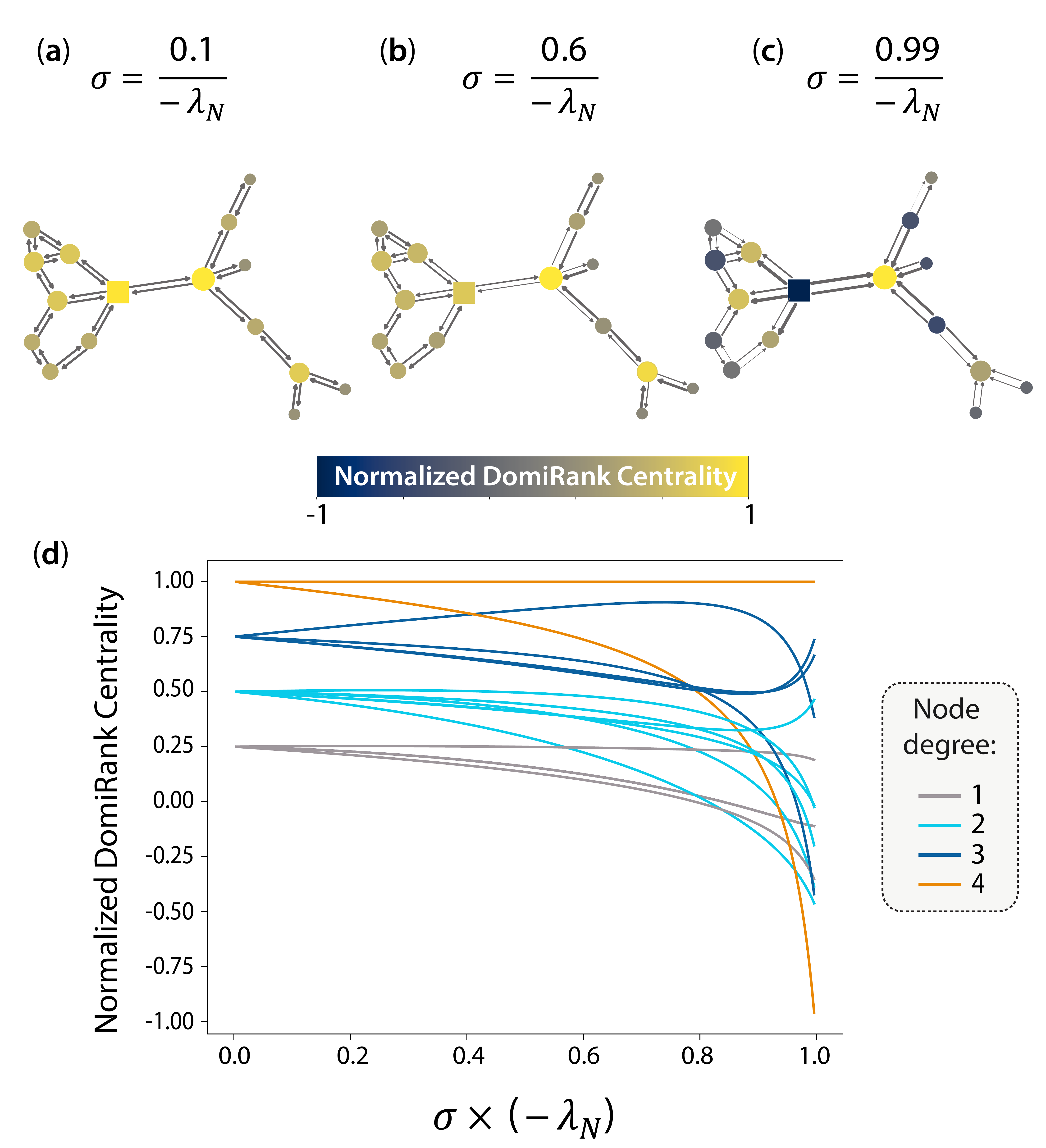}
    \caption{\textbf{DomiRank for different levels of competition ($\sigma$).} DomiRank centrality is displayed on the nodes of a simple network with $N=15$ nodes for (a) low, (b) medium, and (c) large values of $\sigma$. Panel d shows the DomiRank centrality as a function of $\sigma$, wherein each solid line represents a specific node (color encoding node degree). In panels (a-c), the direction of the pairwise transfer of fitness between nodes is shown by arrows, with their thickness representing the magnitude of that exchange. Note that for visualization purposes,  the arrow thickness in panels (a), (b), and (c) are scaled $25:5:1$. }
    \label{fig:phaseSpace}
\end{figure}

The DomiRank centrality is thus modulated by the ratio $\sigma = \frac{\alpha}{\beta}$. To provide further insight into the effect of this parameter on the scores of the centrality, we explore DomiRank for varying values of  $\sigma$ computed for a very simple network (see Fig. \ref{fig:phaseSpace}).   As $\sigma \to 0$, the competition between the different nodes vanishes, and the importance of the nodes reduces to their degree (see Figure \ref{fig:phaseSpace}a,d and Eq. \ref{eq:analytic}). At the other end of the spectrum where $\sigma \to \frac{1}{-\lambda_N}$, the competition is maximum, and although the number of neighbors still plays a role, the network structure is the key feature defining the scores, where the synergistic competitive action of not directly connected nodes might result in their joint dominance in their respective neighborhoods. On that note, Figure \ref{fig:phaseSpace}c,d shows how a node with a relatively high degree (square node) results in the lowest value of DomiRank centrality. This low value is the result of the joint domination by its four neighbors, which, despite having the same or lower degree as the dominated node, are able to increase their relative fitness by dominating their respective non-overlapping unfit neighborhoods and, together, the mentioned node. In fact, at the limit of high $\sigma$, each node tends to be either dominating its neighbor(s) or dominated by its neighbor(s) (see Fig. \ref{fig:phaseSpace}c). This effect is even more apparent when the direction of the steady-state pairwise contribution of the competition term to DomiRank is represented by arrows in Fig.  \ref{fig:phaseSpace}c. Interestingly, in these extremely competitive environments, negative DomiRank scores emerge (see Fig. \ref{fig:phaseSpace}d). Individuals with deficit values of DomiRank are interpreted as fully submissive individuals who, instead of competing, directly give up their resources to neighboring nodes. This is the case for the node highlighted by its square shape in  Fig \ref{fig:phaseSpace}), which for highly competitive environments (Fig. \ref{fig:phaseSpace}c) experiences a reversal of its fitness exchanges (represented by the arrows) when compared with less competitive environments (Fig. \ref{fig:phaseSpace}a). Note that nodes exhibiting negative scores are able to maintain their steady-state DomiRank value due to the relaxation mechanism present in the model (see Eq. \ref{eq:intuition}). This mechanism effectively functions as a recovery or healing process for such nodes.  Finally, intermediate values of $\sigma$ (e.g., see Fig. \ref{fig:phaseSpace}b) represent different domination strategies based on utilizing different balances of local node-based  (low $\sigma$) and global network-structure-based (high $\sigma$) properties.

\begin{figure}[t]
    \centering
    \includegraphics[width = \linewidth]{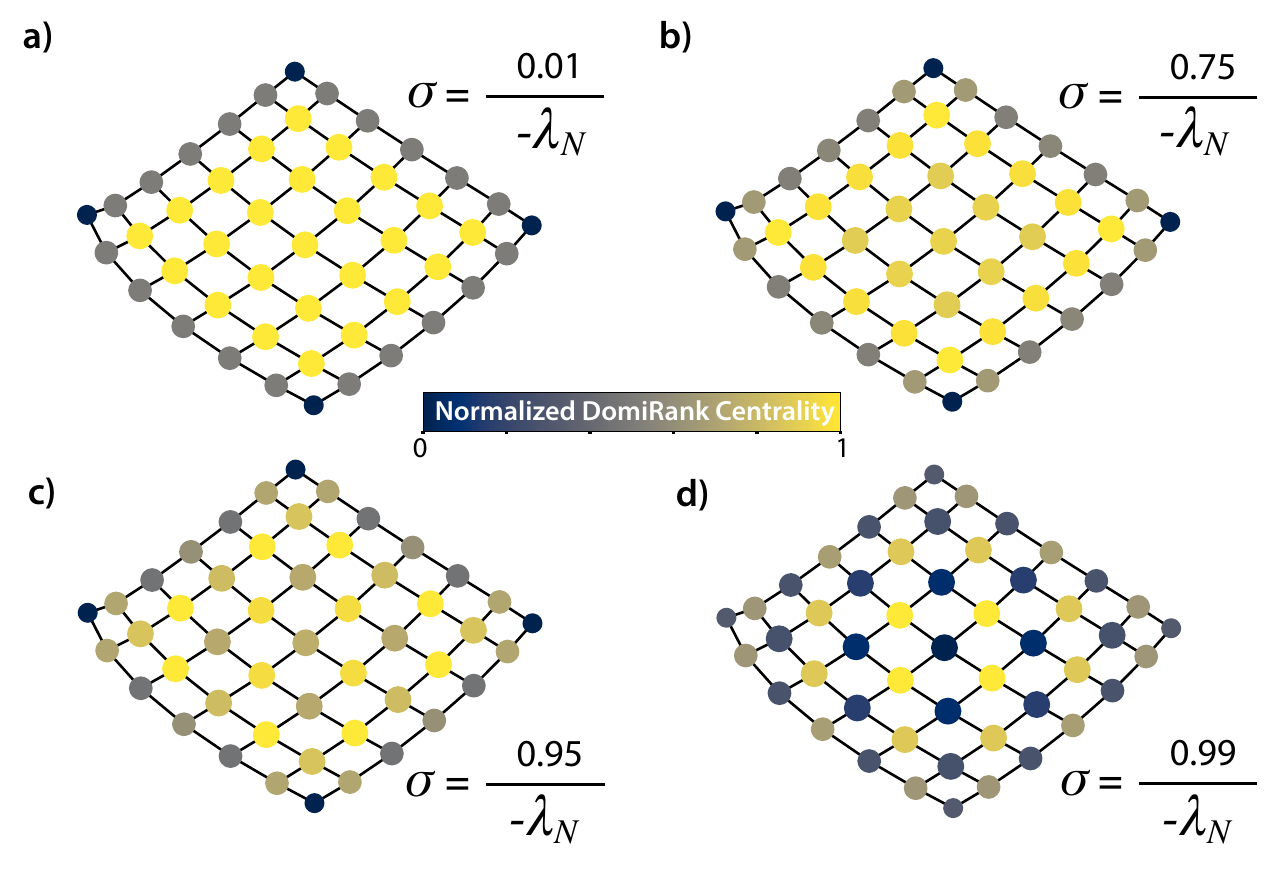}
    \caption{\textbf{The role of $\sigma$ in setting DomiRank values.} DomiRank centrality is displayed on the nodes of a 2D-square lattice with $N=49$ nodes for different values of $\sigma$ (a) $\frac{0.01}{-\lambda_N}$, (b) $\frac{0.75}{-\lambda_N}$, (c) $\frac{0.95}{-\lambda_N}$ and (d) $\frac{0.99}{-\lambda_N}$ to illustrate different levels of competition, and how those levels set the trade-off between local (nodal) and global (meso- to large- scale structure) network information conveyed by DomiRank. Note that for each panel, the values of DomiRank are normalized to range in interval $[0,1]$ for enhanced visualization. }
    \label{fig:LocalGlobal}
\end{figure}

 To develop some intuitive understanding of the role of $\sigma$ in setting up the trade-off between local (nodal) vs. global (meso- to large-scale) network properties in the resulting DomiRank distribution, we explore the DomiRank scores for different values of $\sigma$ in network with clear structure at the mesoscale level, a  square lattice network (see Fig \ref{fig:LocalGlobal}). Note that we have chosen a small domain, $7 \times 7$ (49 nodes) to facilitate the visual interpretation of the results. When $\sigma$ approaches its lower bound (Fig. \ref{fig:LocalGlobal}a), DomiRank converges to the node degree, relying solely on local information. As a result, all nodes except those on the lattice's edges tend to have nearly identical DomiRank values. As $\sigma$ increases, the DomiRank values begin to deviate from the node degree scores (Fig. \ref{fig:LocalGlobal}b). This deviation occurs because each node's DomiRank becomes influenced by the values of their immediate neighbors. As a result, nodes directly linked to the lattice's edge nodes can partially dominate them, increasing their own DomiRank scores (note that this effect appears for significantly smaller values of $\sigma$ than the one displayed in Fig.\ref{fig:LocalGlobal}b but its visualization is less apparent using a consistent color scheme across panels). With further increments in $\sigma$ (Fig. \ref{fig:LocalGlobal}c), the competition dynamics intensify. More internal nodes start sensing the lattice's boundary, and this influence propagates through the DomiRank scores, causing them to adapt based on dominance relationships. In essence, a node's DomiRank score is no longer solely determined by its immediate neighboring nodes; it also considers more distant features. Upon reaching the maximum $\sigma$ value (Fig. \ref{fig:LocalGlobal}d), each node's DomiRank score is partially influenced by the entire network via the competition mechanism. For this extremely competitive setting, an ultimate alternating pattern of dominating and dominated nodes emerges shaped by two global network properties: the finite boundary and global symmetries. Thus, for example, in a square lattice with an even number of nodes, the pattern that emerges differs from the one shown in Fig. \ref{fig:LocalGlobal} due to the different constraints exerted by the system symmetry (for more details, see section S-II in the SM). Also, as an end member, a lattice with periodic boundary conditions, or an infinite lattice, all nodes are indistinguishable from any centrality metric perspective, including DomiRank.

\subsection{Dominance and network fragility}

We also advocate for the capacity of DomiRank to reveal network fragility, both in terms of structure and dynamics. The rationale of this claim ties back to the two mentioned key factors dictating the DomiRank score of a node: its degree (number of neighbors) and the characteristics of its neighbors' neighborhoods (peripheries). In this context, nodes with high degrees that are connected to neighbors with few connections, i.e., sparse peripheries,  are the prime candidates to achieve high DomiRank scores (being a star network, the end-member case of such a configuration).  Those nodes are also central to network fragility, as their failure would lead to the fragmentation of their neighborhood. Interestingly, there exists an alternative source of heightened dominance that relies less on a node's local properties (degree) and more on its position within the global network structure. This kind of dominance primarily emerges in highly competitive environments characterized by high $\sigma$ values. Specifically, it results from joint dominance, where a group of nodes shares an overlapping neighborhood and lacks direct connections among themselves. Consequently, each of these nodes contributes to subduing dominance within the shared neighborhood. Notably, this mechanism also serves to identify vulnerable parts (structures) of the network, as the joint removal of the dominant nodes would lead to the fragmentation of their shared neighborhood, highlighting the fragility inherent in this structure. 

Thus, DomiRank centrality-based attacks are anticipated to be very effective in dismantling networks because of their capacity to shatter the network in small components by targeting preferentially fragile neighborhoods. Also, from the point of view of the dynamics on networks, removing high-score DomiRank nodes could drastically disturb such dynamics, as those removals would appear as insuperable obstructions in sections of the system.

\subsection{Computational Cost}
\begin{figure}[t]
    \centering
    \includegraphics[width = \linewidth]{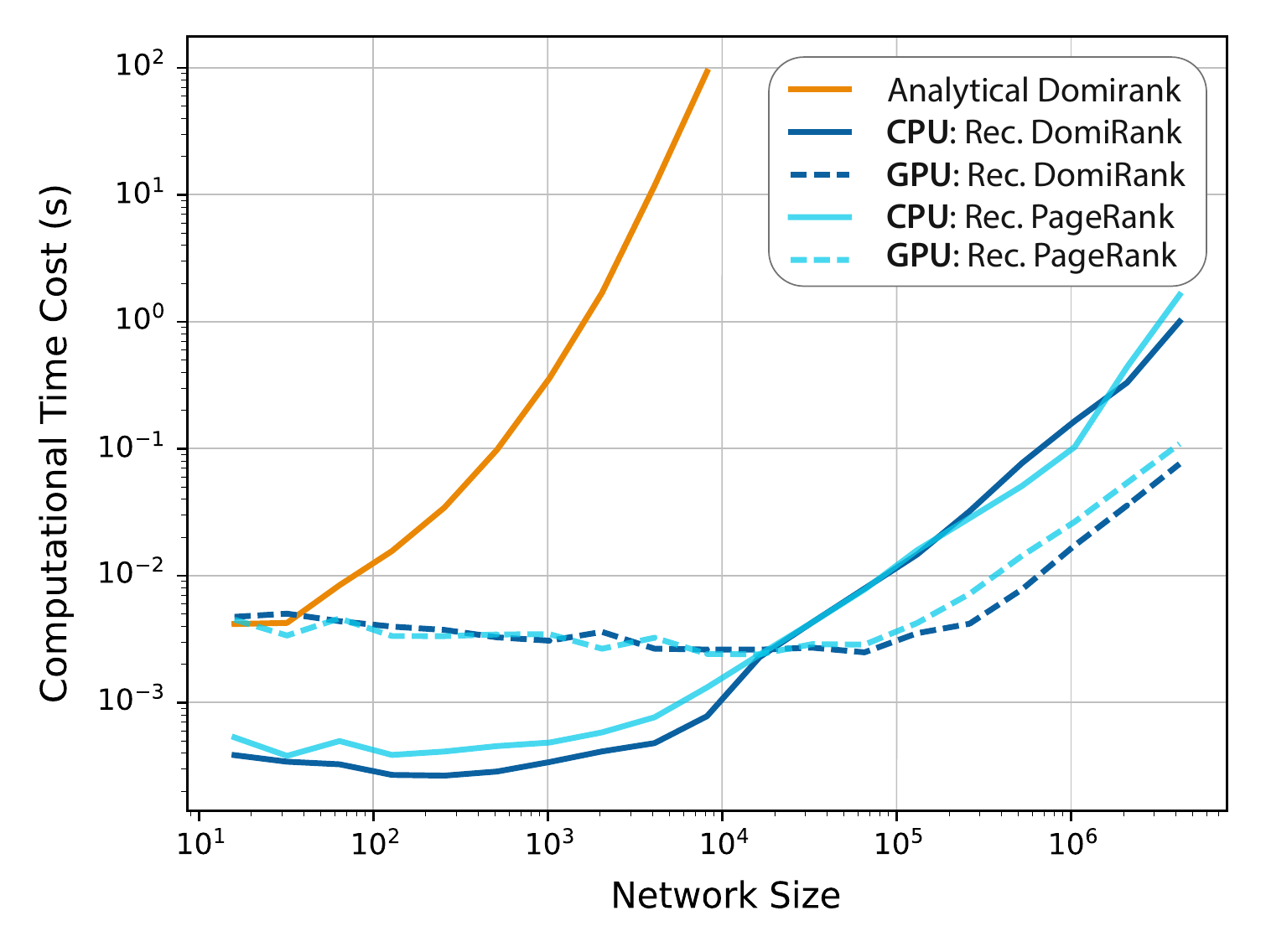}
    \caption{\textbf{Computational cost of DomiRank.} Mean ($30$ samples) computational costs to compute DomiRank analytically (black solid line) and estimate it recursively on a multi-threaded CPU and on the GPU as a function of the network size $N$. The mean DomiRank computational cost is also compared with the mean computational cost for estimating PageRank on the same multi-threaded CPU and GPU.  The convergence criterion is evaluated using the $L1$ error between two consecutive iterations - i.e., $\frac{1}{N}||\boldsymbol{\Gamma}(t) - \boldsymbol{\Gamma}(t+dt)||_1<dt\cdot \epsilon$, with a threshold set to $ \epsilon =  10^{-6}$ (note that for the chosen convergence threshold, the Spearman correlation with the analytical solution is $>0.9999999$).
    }
    \label{fig:ComputationalCost}
\end{figure}

 Beyond the interpretability of DomiRank and its versatility via its free parameter $\sigma$, one of the key advantages of the proposed centrality is that it can be calculated efficiently through iteration in a parallelizable algorithm (see Fig. \ref{fig:ComputationalCost}),
\begin{equation}
\boldsymbol{\Gamma}(t+dt) = \boldsymbol{\Gamma}(t) + \beta [ \sigma A(\boldsymbol{1}_{N\times 1} - \boldsymbol{\Gamma}(t)) - \boldsymbol{\Gamma}(t)]dt,
\label{eq:gpuequation}
\end{equation}
with a computational cost $C$:%
\begin{equation}
C(t,A) = t(m + 5N),
\label{eq:compcost}
\end{equation}
which scales with $\mathcal{O}(m+N)$, where $m$ is the number of links and $N$ is the number of nodes. Thus, the DomiRank scales with $\mathcal{O}(N^2)$ in the worst case (fully connected graph). Importantly, the algorithm can be distributed among $\kappa$ cores given that $\kappa \leq m$ for sparse matrices, which allows for parallel computation and efficient execution on GPUs. Fig. \ref{fig:ComputationalCost} shows the computational costs of calculating DomiRank (analytically and recursively)  and PageRank (recursively) for different network sizes, showing: (i) the high computational cost for the analytic computation of DomiRank, as it requires matrix inversion, (ii) the comparable computational cost of the DomiRank to that of PageRank on both CPU and GPU infrastructure, and (iii) that the latency of computing DomiRank on the GPU is the computational bottleneck unless the number of links is significantly larger the number of GPU cores, i.e.,  $m>>\kappa$. Thus, DomiRank centrality is computable even for massive (sparse) networks, allowing computational time costs under one second for networks consisting of millions of nodes.

\section{Evaluating DomiRank} 

In order to gain further insight into the capabilities of DomiRank and to benchmark its performance with respect to the other most commonly used centralities, we examine the efficacy of targeted attacks based on DomiRank centrality for different network topologies, analyzing its ability to dismantle the network structure and functionality, and contrasting its performance with those of the attacks based on other centralities.

\begin{figure*}
    \centering
    \includegraphics[width = \linewidth]{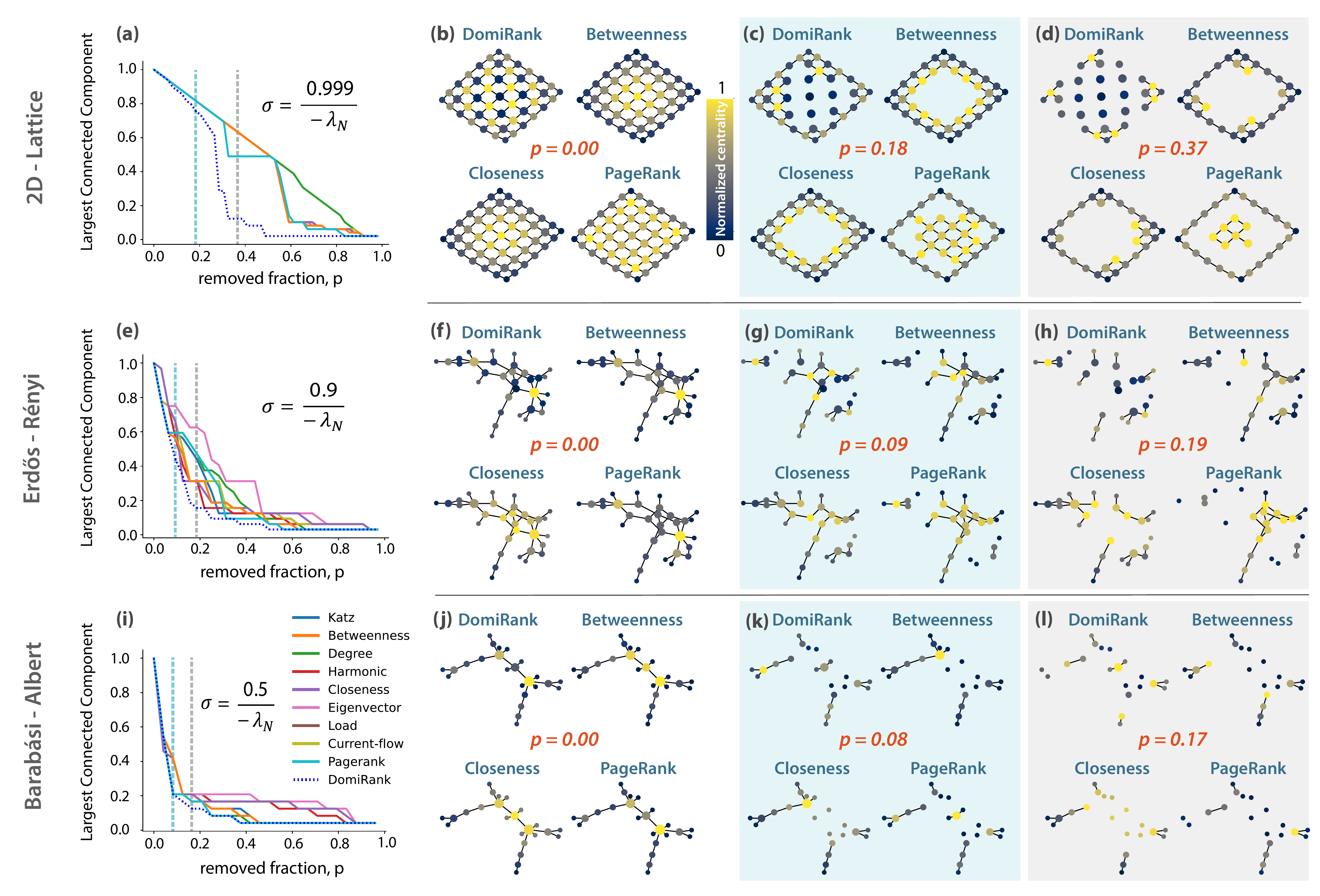}
    \caption{\textbf{Comparing DomiRank with other centralities in dismantling toy networks.} Evolution of the relative size of the largest connected component whilst undergoing sequential node removal according to descending scores of various centralities for three toy networks: (a) 2D regular lattice $(N=49)$, (e) Erd\H{o}s-R\'enyi (ER;  $N=32$),  and (i) Barab{\'a}si-Albert (BA; $N=25$). For each topology, panels (b-d), (f-h), and (j-l) show the graphical representation of the respective networks at various stages of the attack based on DomiRank, betweenness, closeness, and PageRank centralities. Note that the nodes are colored according to the relative value of the centralities normalized to be in an interval $[0,1]$ for enhancing comparability and visualization purposes.}
    \label{fig:smallScale}
\end{figure*}

\subsection{Structural Network robustness}

In this section, we evaluate the structural robustness of different networks, both synthetic and real-world topologies, under sequential node removal (attacks) based on different centrality metrics and compare the results with those obtained based on DomiRank.  To evaluate network robustness, we use its most commonly used proxy,  the evolution of the relative size of the largest connected component ($LCC$) \cite{Trajanovski2013, Ventresca2015, Williams2016, Cats2020}, whilst the network is undergoing sequential node removal. We compare the robustness of the different networks for the different attacks by directly comparing the resulting $LCC$ curves, and for simplicity and enhanced comparability, we also use the area under that curve as a summary indicator of robustness (the larger the area, the more robust the network is under that particular attack).

We start our analysis with synthetic toy networks, consisting of a reduced number of nodes, but wherein their graphical representation still allows us to visually identify patterns on the centrality distributions for different topologies, gaining insight into the interpretation of DomiRank and its performance when compared with different centralities. Particularly, we perform targeted attacks based on DomiRank and nine other centralities for three different topologies:  2D-regular lattice \cite{biggs1993algebraic}, Erd\H{o}s-R\'enyi \cite{Erdos1959}, and a Barab{\'a}si-Albert \cite{Barabasi1999} networks. Note that for each topology, the range of $\sigma$ was explored to determine its optimal value to dismantle the network, i.e., minimize area under the $LCC$ curve (for other centralities such as Katz and PageRank, we used, without loss of generality of the results, the default values of 0.01 and 0.85, respectively  - See SM section S-III for details). Fig. \ref{fig:smallScale}a,e,i reveals that the DomiRank centrality-based attacks dismantle these three networks more efficiently than all other tested centrality-based attacks. More particularly, DomiRank excels at dismantling regular networks (Fig. \ref{fig:smallScale}a). It is not surprising that for this topology, DomiRank centrality produces the most effective attack for large values of $\sigma$, wherein network structure is overweighed to the detriment of local node properties. This value of $\sigma$ leads to a DomiRank distribution wherein if a node is important (\textit{dominating} node), all of its adjacent nodes are not important (\textit{dominated} node), and vice-versa (Fig \ref{fig:smallScale}b). An attack strategy based on such an alternating spatial pattern is significantly advantageous with respect to other traditional centrality-based attacks (Fig \ref{fig:smallScale}a-d). This advantage stems from the strategic removal of existing neighbors, effectively isolating nodes and efficiently reducing the size of the largest connected component ($LCC$). Applying a similar DomiRank-based attack strategy to a more heterogeneous network, such as Erd\H{o}s-R\'enyi (see Fig. \ref{fig:smallScale}e), still leads to the highest fragility of the network, also capitalizing on the built-in competition mechanism of DomiRank (high value of $\sigma$) that penalizes connections between nodes labeled as highly central, reducing the situations wherein connected nodes possess disjoint neighborhoods to exert their respective dominance. For most of the other centrality metrics, including Betweeness, Eigenvector, PageRank, and Katz, highly central nodes permeate their centrality to their direct connections (see Fig. \ref{fig:smallScale}f-h). However, that {\it by-contact} importance only reflects the centrality of their {\it truly} important neighbor, yielding attack sequences less efficient than DomiRank.  As networks display more hub-dominated topologies (e.g., scale-free), we expect that the optimal value of $\sigma$ for the most efficient attack decreases, emphasizing nodal properties (degree) with respect to the neighborhood structure. In the toy example for a network generated by a Barab{\'a}si-Albert model (see Fig. \ref{fig:smallScale}i), DomiRank still outperforms other centrality-based attacks in dismantling the network. In this case, the improvement is incremental since the most relevant information to destroy the network is local (node degree), and most of the centralities converge to a similar nodal ranking and attack sequence (Fig. \ref{fig:smallScale}j-l).   

\begin{figure*}
    \centering
    \includegraphics[width = \linewidth]{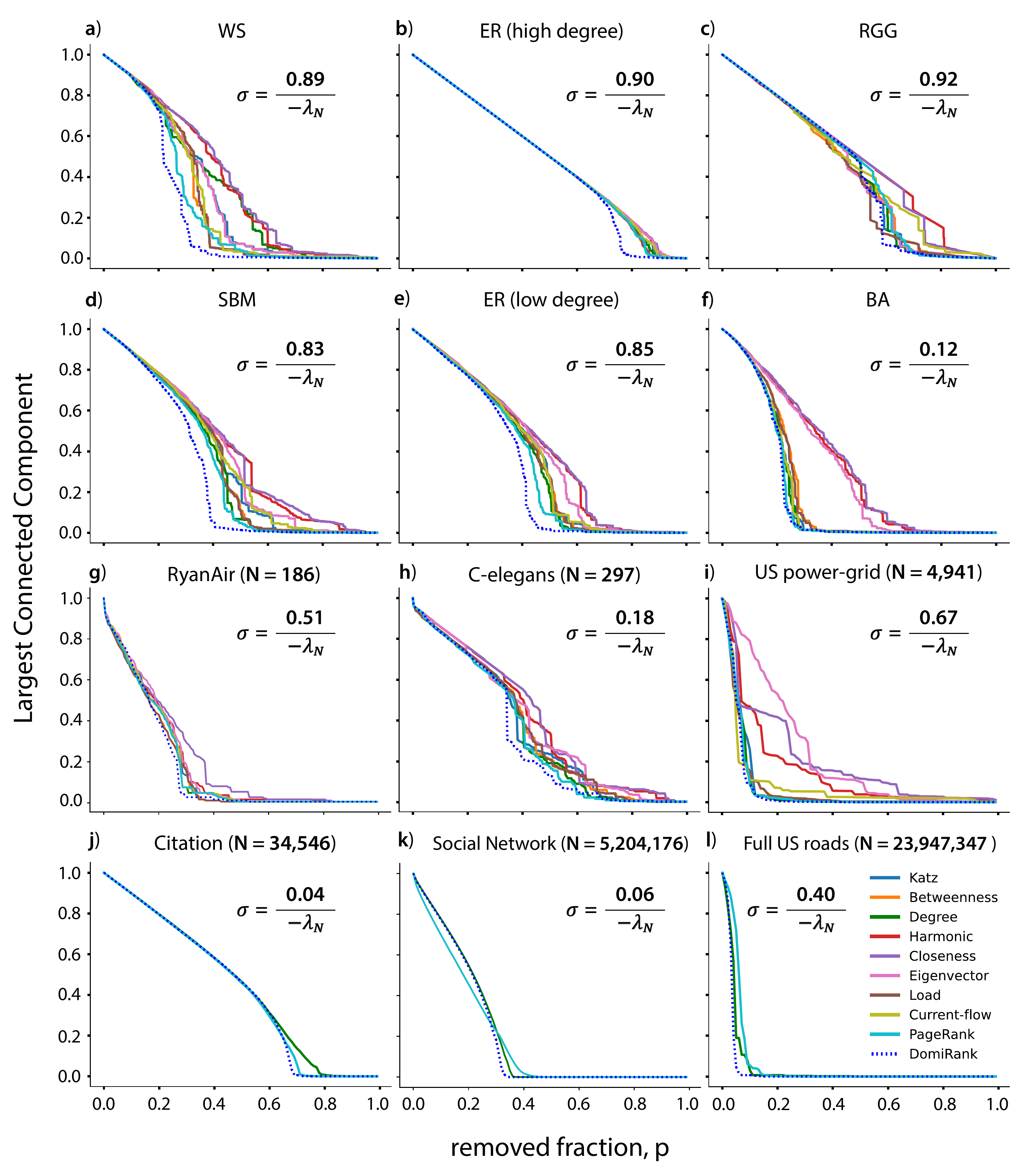}
    \caption{\textbf{Centrality-based attacks on synthetic and real-world networks.} Evolution of the relative size of the largest connected component (robustness) whilst undergoing sequential node removal according to descending scores of various centrality measures for different synthetic networks of size $N=1000$: (a) Watts-Strogratz (WS; small-world, $\bar{k} = 4$), Erd\H{o}s-R\'enyi (ER) with (b) high ($\bar{k} = 20$) and (e) low link density ($\bar{k} = 6$), (c) random geometric graph (RGG; $\bar{k} = 16$), (d) stochastic block model (SBM; $\bar{k} = 7$), and (f) Barab{\'a}si-Albert (BA; $\bar{k} = 6$). The performance of the attacks based on the different centrality metrics is also shown for different real networks: (g) hub-dominated transport network (airline connections, $\bar{k} = 16$), (h) neural network (worm, $\bar{k} = 29$), (i) spatial network (power-grid, $\bar{k} = 3$), (j) citation network ($\bar{k} = 25$), (k) massive social network ($\bar{k} = 19$), and (l) massive spatial transport network (roads, $\bar{k} = 5$). Note that for panels j, k, and l, where massive networks are shown, only a few attack strategies are displayed due to the impossibility of computation of the rest. }
    \label{fig:SyntheticReal_Networks}
\end{figure*}

We further investigate the efficacy of the attack strategies based on the DomiRank centrality by dismantling larger synthetic networks ($N=1000$) with varying degrees ($2 < \bar{k} < 20$) for numerous topologies. Particularly, we analyze the robustness of Watts-Strogatz \cite{Watts1998}, stochastic-block-model \cite{holland1983stochastic}, Erd\H{o}s-R\'enyi,  random geometric graph \cite{gilbert1959random}, and Barab{\'a}si-Albert networks, under ten different targeted attack strategies based on different centralities, including DomiRank, which revealed itself as the overall most efficient at dismantling synthetic networks (Fig. \ref{fig:SyntheticReal_Networks}a-f). As hinted from our previous analysis of the toy networks, the margin by which the DomiRank-based attack outperforms the other strategies relates to the topological properties of the networks, which also dictate the optimal value of $\sigma$. Thus, for the Barab{\'a}si-Albert topology (hub-dominated), DomiRank offers only an incremental improvement in the efficiency at dismantling the network (see Fig. \ref{fig:SyntheticReal_Networks}f). On the other hand, for networks with meso-to-macro scale structural features (e.g., regularity) that dominate over the local node-based properties, DomiRank centrality significantly outperforms all other centralities (Fig. \ref{fig:SyntheticReal_Networks}d). This also occurs for the Erd\H{o}s-R\'enyi (Fig. \ref{fig:SyntheticReal_Networks}b,e) and Watts-Strogatz networks (Fig. \ref{fig:SyntheticReal_Networks}a). For a more detailed comparison between the different centralities, we refer the reader to the SM (section S-IV) where the correlation between them is displayed.

\begin{figure*}
    \centering
    \includegraphics[width = \linewidth]{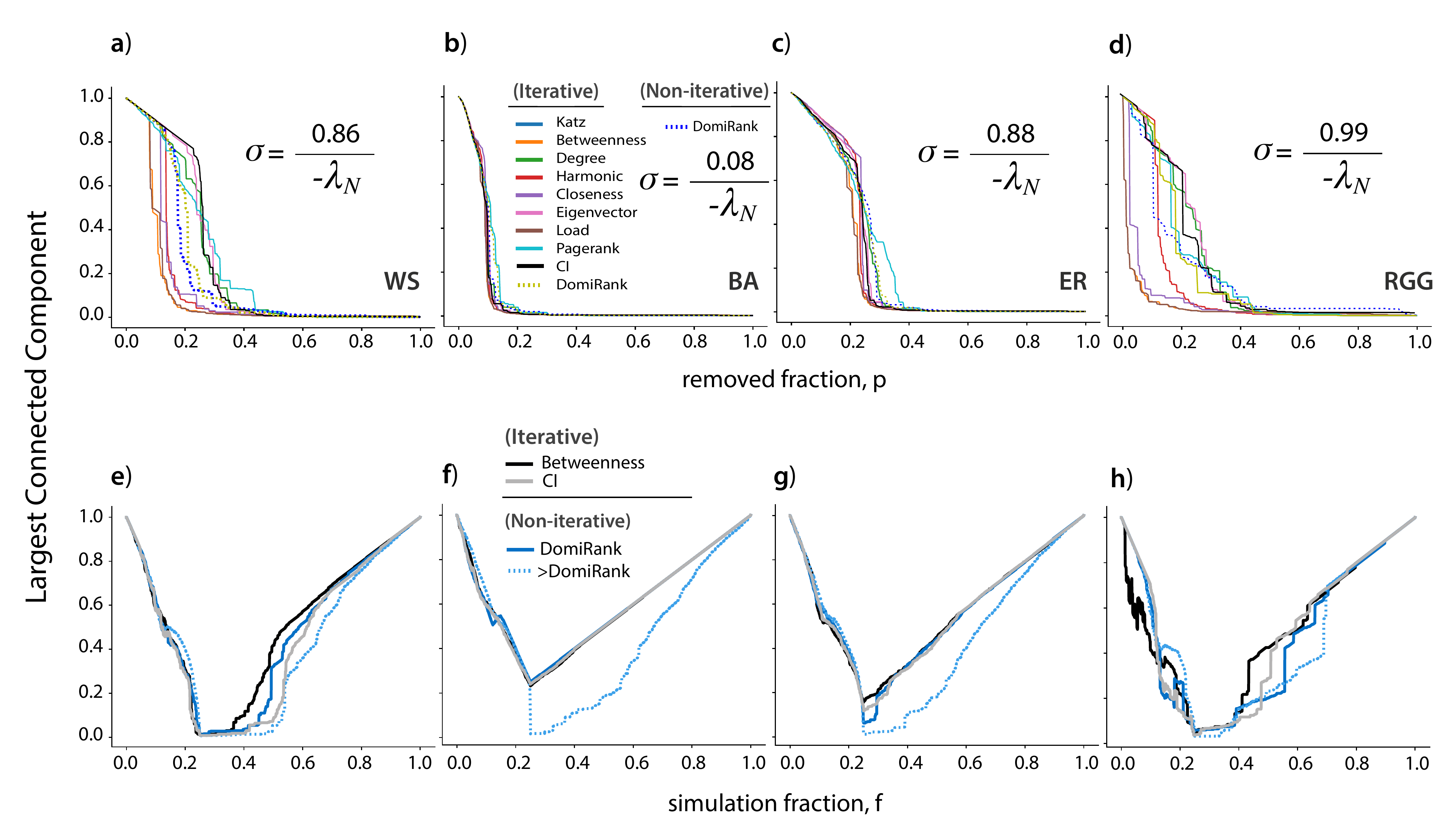}
    \caption{\textbf{Assessing the effect of iterative centrality-based attacks and recovery mechanisms on network resilience.} Panels a-d show the evolution of the relative size of the largest connected component of various synthetic networks of size $N=500$, namely; (a) Watts-Strogatz (WS; $\bar{k}=4$), (b) Barab{\'a}si-Albert (BA; $\bar{k}=4$), (c) Erd\H{o}s-R\'enyi (ER; $\bar{k}=4$), and (d) random geometric graph (RGG; $\bar{k}=7$), undergoing sequential node removal based on iteratively computed centralities and on pre-computed DomiRank. Panels e-h show the evolution of the relative size of the largest connected component for the same networks undergoing sequential node removal based on pre-computed DomiRank (optimal and high $\sigma$), iterative betweenness, and Collective Influence (CI), where a stochastic first-in-first-out node recovery  (stack recovery implementation) process with a probability of recovery $p=0.25$ each time step is implemented. 
    }
    \label{fig:iterative}
\end{figure*}

Real networks introduce several properties that are hard to produce simultaneously using generative models. Therefore for a more thorough and general benchmark of DomiRank, we analyze various real networks topologies of various sizes: (g) hub-dominated transport network (RyanAir connections) \cite{Tejedor2017_AI}, (h) neural network (C-elegans) \cite{manke2006entropic, jeong2001lethality, nr}, (i) spatial network (power-grid of the Western States of the United States of America) \cite{Watts1998}, (j) citation network (high-energy-physics arXiv) \cite{kunegis2013konect, leskovec2007graph}, (k) massive social network (LiveJournal users and their connections) \cite{leskovec2007graph}, and (l) massive spatial transport network (Full US roads) \cite{kunegis2013konect}. Our results are in line with those for synthetic networks, showing that the DomiRank is able to consistently dismantle the networks more efficiently than the other centrality-based attacks tested (see Fig. \ref{fig:SyntheticReal_Networks}g-l). Another interesting phenomenon also observed for the synthetic networks is that the DomiRank-based attacks remove links more efficiently than previous methods (see Section S-VII in the SM). This means that for many of these networks, not only is the DomiRank better at reducing the size of the largest cluster size, but it also more severely cripples its connectivity, yielding not only to an overall faster but also a more thorough dismantling of the network. However, we note that for the social network analyzed (Fig \ref{fig:SyntheticReal_Networks}k), the PageRank-based attack outperforms the one based on DomiRank. We attribute this phenomenon to the presence of structural heterogeneity in the network topology (i.e., different structures in different subgraphs of the network). This heterogeneity hinders the assessment of node importance by DomiRank with a single value of $\sigma$ for the whole network. In the SM (section S-V), we provide evidence showing that, indeed, heterogeneity can lead DomiRank to underperform, hinting at potential approaches to address the evaluation of networks exhibiting heterogeneity.  

As a last point of discussion about Fig \ref{fig:SyntheticReal_Networks}, we want to emphasize how the sensitivity of DomiRank to its unique parameter $\sigma$ becomes apparent from analyzing and displaying the results for such an extensive set of diverse networks jointly. This sensitivity, far from being a weakness, is a key strength of DomiRank, as it allows us to assess the important nodes in networks with topologies as different as a regular lattice and hub-dominated network.  Thus, for any individual case, the edge that a DomiRank-based attack could offer over other centrality-based attacks could vary from being very significant (e.g., planar networks) to just marginal (e.g., scale-free). In fact, it is the consistent and sustained superior performance of DomiRank-based attacks across all ranges of topologies that makes these results particularly noteworthy overall.    

The analysis of synthetic networks and real-world topologies has demonstrated the capacity of DomiRank to integrate local (node) and mesoscale information of the network, which, together with the competition mechanism embedded in its definition, produces centrality distributions that efficiently dismantle the networks by avoiding redundant scores in neighboring nodes (importance by-contact). This apparent handicap for other centralities could be addressed at the cost of recomputing the centrality distributions after each node removal. Note that this cost is prohibitive for distance-based or process-based metrics such as closeness, betweenness, or load centralities, even for networks of modest sizes as, for instance, betweenness has a computational complexity that scales with $\mathcal{O}(Nm)$ and $\mathcal{O}(Nm + N^2\log N)$ for unweighted and weighted graphs respectively \cite{brandes2001faster}. Despite this limitation, and for the sake of completeness, we also benchmark the  DomiRank centrality distribution (computed once before the beginning of the attack) with the targeted attacks based on sequentially recomputed centralities.  In this part of the analysis, we have incorporated attacks based on the Collective Influence (CI) algorithm \cite{morone2015influence, morone2016collective, Pei2018}, which is an iteratively recomputed centrality that aims to find the most influential nodes in a network. CI could be particularly relevant to our study as: (i)  it can be mapped to an optimal percolation problem, and (ii) it has been successful in identifying previously neglected (non-locally important) nodes as important influencers. 

\begin{figure*}
    \centering
    \includegraphics[width = \linewidth]{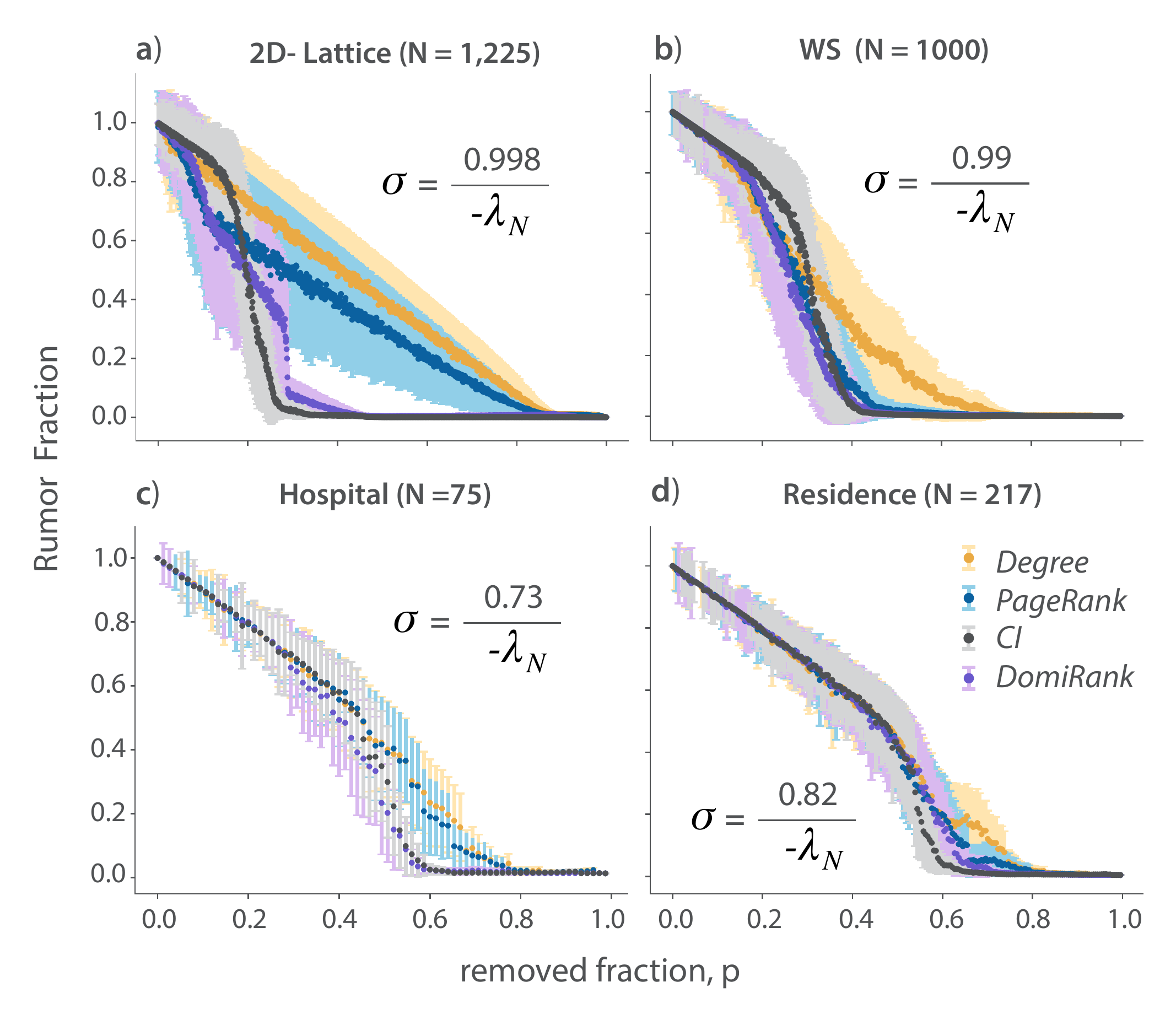}
    \caption{\textbf{Functional robustness of synthetic and real-world networks under centrality-based attacks.} Average rumor spread fraction (error-bars representing the standard deviation) of 1000 rumor spreading simulations as a function of the subsequent network stage resulting of sequential node removal according to degree, PageRank, DomiRank, and Collective Influence strategies, for two synthetic networks: (a) 2D regular lattice ($\bar{k}=4$) and (b) Watts-Strogatz (WS; $\bar{k}=6$), and two real networks: (c) a contact-tracing social network 
    (Hospital; $\bar{k}=30$) and (d) a survey based social network 
    (Residence; $\bar{k}=16$).} 
    \label{fig:functionalRobustness}
\end{figure*}

Fig. \ref{fig:iterative}a-d displays the increase in performance of various centrality-based attacks when recomputed iteratively, particularly for betweenness centrality. In fact, for all the synthetic topologies tested, iterative betweenness and load centralities lead to the most efficient attacks at dismantling networks by a large margin. We also note, that CI is able to fully disintegrate the network (i.e., reduce all clusters to minimum size) the fastest, as expected from its mathematical formulation. However, the path of deterioration followed by the $LCC$ during the attack is not among the most competitive in rapidly and sustainably reducing the size of the largest connected component. Notably, the attacks based on pre-computed DomiRank centrality generally outperform other attacks based on iterative centralities that are computationally feasible for medium, large, and massive networks - i.e., iterative degree, PageRank, eigenvector, Katz, and CI (except for ER and marginally for BA). Note that attacks based on iterative DomiRank centrality perform worse than the ones obtained from a single computation, which is actually expected as DomiRank leads to attack strategies aiming to cause structural damage, which requires the joint removal of several nodes. Therefore by recomputing DomiRank every time step, no coherent strategy emerges as the network structure becomes a moving target, i.e., the structure is re-evaluated at a faster rate (every node removal) than the time needed to remove the number of nodes necessary to inflict the structural damage. This fact underscores the intrinsic ability of the pre-computed DomiRank version to extract local information while considering the network's global context (with a larger value of $\sigma$ indicating a broader global context). In other words, DomiRank's underlying competition mechanism establishes an inherent, built-in process to prevent the assignment of artificial redundant scores to neighboring nodes. All the other displayed methods in Fig. \ref{fig:iterative}a-d lack this intrinsic mechanism, which is compensated for through the sequential recomputation of the centralities after each node removal. 

From all the attacks shown in Fig. \ref{fig:iterative}, we want to highlight iterative betweenness, being the most efficient, and CI because it shares potential similarities with DomiRank. Attacks based on iterative betweenness centrality excel at destroying the $LCC$ by finding {\it bottle-neck} nodes instrumental in mediating most of the shortest paths and, thus, focusing on simply splitting the network. As a result of these fundamental differences in the aim of the two centralities, we expect that despite the DomiRank-based attack being less efficient at dismantling the network than those based on iterative betweenness, it causes more severe and enduring damage, making it more difficult to recover from when compared with the damage produced by an iterative betweenness attack. CI-based attacks seek to find influential nodes according to their potential to cascade down information. As such, CI does not look only to local information but also integrates information regarding the neighboring structure of the nodes. Despite both CI and DomiRank having the ability to combine local and global network properties, they differ fundamentally in their mechanism to do it, and therefore, their nodal importance assessment could differ significantly (See section S-VI in the SM for more details). In fact, we expect DomiRank (particularly for large $\sigma$) to generally inflict more enduring damage than CI since DomiRank focuses more on fragile neighborhoods, whose fragmentation depends on joint sets of nodes, and once removed, the restitution of a fraction of those nodes might not serve to recover a proportionally equivalent fragmented section. 

The first indirect piece of evidence supporting the hypothesis that DomiRank inflicts more severe and enduring damage than other centrality-based attacks is that DomiRank-based attacks remove links more efficiently than other attack strategies (see section S-VII in the SM). To test the hypothesis more directly, we implemented two simple recovery mechanisms to evaluate from which of the attacks the network was less prompt to recover. Both recovery mechanisms assign a probability $p$ to a given removed node to recover every time step, wherein the first strategy selects the nodes in the same order that they were removed (results shown in Fig. \ref{fig:iterative}e-h), while for the second strategy, nodes are selected at random from the pool of removed nodes (see results in section S-VIII in the SM). Our results show that for all the networks, except for the random geometric graph (probably due to network modularity), when a recovery mechanism is put in place, the attack based on a single computation of DomiRank centrality has a comparative dismantling ability than the attack based on iterative betweenness, as shown by the deterioration trend of the $LCC$ in Fig. \ref{fig:iterative}e-h. Moreover, for all the analyzed topologies, the DomiRank-based attack causes longer-lasting effects, as the recovery mechanism requires a larger fraction of reinstated nodes to obtain an equivalent recovery in terms of $LCC$.  The superior ability of the DomiRank strategy to inflict more severe damage is grounded in its aim to dismantle the inherent network structure via the dominance mechanism. To further demonstrate this point,  Fig. \ref{fig:iterative}e-h also displays a high-$\sigma$ DomiRank-based attack (boosted dominance), where the pace at which the networks recovered was increasingly impeded. Thus, the DomiRank centrality provides a trade-off between the celerity and the severity of the attack through modulation of $\sigma$, highlighting its applicability to design vaccination schemes and other mitigation strategies.

\subsection{Functional Network robustness}
Sequential node failure caused by random or targeted attacks can compromise not only the structure but also the dynamics taking place on the network, i.e., the functional robustness of the network. In this section, we benchmark the ability of DomiRank-based attacks to disrupt a rumor-spreading dynamic \cite{moreno2004dynamics} on different network topologies.  We implement an epidemic-like model for spreading rumors, where each node represents an individual who can be in three potential states with respect to the rumor:  ignorant, active spreader, and stifler (have heard the rumor but is no longer spreading it) \cite{zhao2012sihr}. More specifically, the rumor-spreading dynamic takes four arguments: (i) the network $\mathcal{N}$, (ii) the origin of the rumor (node), (iii) the probability of persuading someone of the rumor ($\rho_r$), and (iv) the probability of becoming a stifler ($\rho_s$). We implement this model on the subsequent networks originating from sequences of node removal according to different centrality-based targeted attacks, choosing the fraction of persons that believe the rumor at the end of the process as the proxy for functional robustness. Specifically, we contrast the results obtained from using a DomiRank-based attack with three other relevant centralities: Degree, PageRank, and CI. The selection of these three metrics is based on different reasons. Degree is a simple, widely-used centrality, which at the same time corresponds to the limit of no competition for DomiRank ($\sigma=0$). PageRank is arguably one of the most effective metrics to identify critical nodes, as shown in the results in Fig. \ref{fig:SyntheticReal_Networks}. We also include CI as an example of a sequentially recomputed centrality (after each nodal removal) that is particularly relevant to information-spreading dynamics.

Fig. \ref{fig:functionalRobustness} showcases the superior ability of the DomiRank-based attacks to halt a rumor-spreading process in comparison with degree- and PageRank-based attacks for synthetic and real networks.  When compared with CI, a general tendency emerges: DomiRank outperforms in terms of disrupting the spreading process for smaller fractions of nodes removed, but as more nodes are removed, CI becomes competitive or even outperforms DomiRank (see Fig. \ref{fig:functionalRobustness}). We attribute this to the fact that DomiRank attempts to halt the spreading process by creating obstructions for the spreading process by removing sets of locally dominant nodes in the network, which very effectively impedes the spreading of information in their neighborhoods. In other words, contrasting with the goal of CI to undermine the multiplicative contagion effect by targeting the removal of potential super-spreaders, DomiRank-based attacks aim to stall the spreading process by fragmenting the spreading domain, quasi-isolating neighborhoods in the network.

The ability of DomiRank to highlight the set of nodes to effectively establish firewalls to mitigate the propagation of rumors is conceptually generalizable to other dynamic processes, such as information transport or epidemic spreading, to name a few, prompting the idea that the DomiRank could be used for establishing efficient vaccination schemes.

\section{Concluding Remarks}

This work presents a new centrality metric, called DomiRank, which evaluates nodal importance by integrating different aspects of the network's topology according to a single tunable parameter that controls the trade-off between local (nodal) and mesoscale (structural) information considered. Thus, the competition mechanism embedded in the definition of DomiRank offers a novel perspective on identifying highly important nodes for network functionality and integrity by contextualizing the relevance of nodes in their respective neighborhoods, taking into account emergent synergies between not directly connected nodes over overlapping neighborhoods (i.e., joint dominance).  

One key feature of DomiRank centrality is its low computational cost and fast convergence.  On this front, the DomiRank centrality is competitive with the PageRank centrality whilst being parallelizable, which allows for efficient execution on GPU infrastructure, making it applicable on massive sparse networks. 

We show the superior ability of DomiRank to generate effective targeted attacks to dismantle the network structure and disrupt its functionality, offering an outstanding trade-off between the celerity and the severity of the attack and, therefore, significantly reducing network resilience. DomiRank could be further customized to account for localizing heterogeneity in the topology of massive real-world networks, enhancing the assessment of nodal importance in such systems.
 Also, we anticipate that hybrid attack strategies, where DomiRank is recomputed at different stages of the attack process, might also increase its performance. Moreover, analyzing the robustness of networks in the light of the recently introduced Idle Network (connectivity of the removed nodes by an attack) \cite{Tejedor2017_AI, engsig2022}  could be particularly illuminating as the DomiRank's parameter exerts a direct control on the fragmentation of the Idle network.

Finally, we want to highlight the broad applicability of DomiRank centrality to different domains, as via its versatile dominance mechanism, it is anticipated to be instrumental for tasks as diverse as improving SPAM detection, establishing effective vaccination schemes, or assessing vulnerabilities in transportation networks, just to name a few. Thus, DomiRank, by revealing fundamental aspects of network fragility, can spur further research to develop more effective mitigation strategies to improve our overall understanding of complex systems structure and resilience.

\appendix

\section*{Appendix}

\textbf{Proof:} \textit{We define the dominance centrality $\boldsymbol{\Gamma}$ as the stationary solution of  equation \ref{eq:intuition}:}
\begin{equation}
\frac{1}{\beta}\frac{\partial \boldsymbol{\Gamma}(t)}{\partial t} = \sigma A(\theta\boldsymbol{1}_{N\times 1} - \boldsymbol{\Gamma}(t)) -  \boldsymbol{\Gamma}(t),
\end{equation}
 \textit{By definition, the centrality only exists if $\boldsymbol{\Gamma}(t)$ converges to $\boldsymbol{\Gamma}$, and thus;}
\begin{equation}
\lim_{t\to\infty}\frac{\partial \boldsymbol{\Gamma}(t)}{\partial t} = 0,
\end{equation}
\textit{which implies,}
\begin{equation}
\lim_{t\to\infty}[\sigma A(\theta\boldsymbol{1}_{N\times 1} - \boldsymbol{\Gamma}(t)) -  \boldsymbol{\Gamma}(t)] = 0
\label{eq:convergence}
\end{equation}
\textit{thus, we can solve eq. \ref{eq:convergence} in the following manner;}
\begin{equation}
    \sigma \theta A\boldsymbol{1}_{N\times 1} - \lim_{t\to\infty}[(\sigma A + I_{N\times N})\boldsymbol{\Gamma}(t)] = 0,
\end{equation}

\textit{and therefore,}

\begin{equation}
    \lim_{t\to\infty} \boldsymbol{\Gamma}(t) := \boldsymbol{\Gamma} = \sigma \theta (\sigma A + I_{N\times N})^{-1} A \boldsymbol{1}_{N\times 1},
    \label{eq:outdegree}
\end{equation}

\section*{Data Availability}
The data supporting the findings of this study are available within the paper and references.
\vspace{4mm}
\section*{Code Availability}
The code developed to compute DomiRank centrality is available at 10.5281/zenodo.8369910.


\begin{thebibliography}{57}%
\makeatletter
\providecommand \@ifxundefined [1]{%
 \@ifx{#1\undefined}
}%
\providecommand \@ifnum [1]{%
 \ifnum #1\expandafter \@firstoftwo
 \else \expandafter \@secondoftwo
 \fi
}%
\providecommand \@ifx [1]{%
 \ifx #1\expandafter \@firstoftwo
 \else \expandafter \@secondoftwo
 \fi
}%
\providecommand \natexlab [1]{#1}%
\providecommand \enquote  [1]{``#1''}%
\providecommand \bibnamefont  [1]{#1}%
\providecommand \bibfnamefont [1]{#1}%
\providecommand \citenamefont [1]{#1}%
\providecommand \href@noop [0]{\@secondoftwo}%
\providecommand \href [0]{\begingroup \@sanitize@url \@href}%
\providecommand \@href[1]{\@@startlink{#1}\@@href}%
\providecommand \@@href[1]{\endgroup#1\@@endlink}%
\providecommand \@sanitize@url [0]{\catcode `\\12\catcode `\$12\catcode `\&12\catcode `\#12\catcode `\^12\catcode `\_12\catcode `\%12\relax}%
\providecommand \@@startlink[1]{}%
\providecommand \@@endlink[0]{}%
\providecommand \url  [0]{\begingroup\@sanitize@url \@url }%
\providecommand \@url [1]{\endgroup\@href {#1}{\urlprefix }}%
\providecommand \urlprefix  [0]{URL }%
\providecommand \Eprint [0]{\href }%
\providecommand \doibase [0]{https://doi.org/}%
\providecommand \selectlanguage [0]{\@gobble}%
\providecommand \bibinfo  [0]{\@secondoftwo}%
\providecommand \bibfield  [0]{\@secondoftwo}%
\providecommand \translation [1]{[#1]}%
\providecommand \BibitemOpen [0]{}%
\providecommand \bibitemStop [0]{}%
\providecommand \bibitemNoStop [0]{.\EOS\space}%
\providecommand \EOS [0]{\spacefactor3000\relax}%
\providecommand \BibitemShut  [1]{\csname bibitem#1\endcsname}%
\let\auto@bib@innerbib\@empty
\bibitem [{\citenamefont {Albert}\ \emph {et~al.}(2000)\citenamefont {Albert}, \citenamefont {Jeong},\ and\ \citenamefont {Barab{\'a}si}}]{Albert2000}%
  \BibitemOpen
  \bibfield  {author} {\bibinfo {author} {\bibfnamefont {R.}~\bibnamefont {Albert}}, \bibinfo {author} {\bibfnamefont {H.}~\bibnamefont {Jeong}},\ and\ \bibinfo {author} {\bibfnamefont {A.-L.}\ \bibnamefont {Barab{\'a}si}},\ }\bibfield  {title} {\bibinfo {title} {Error and attack tolerance of complex networks},\ }\href {https://doi.org/10.1038/35019019} {\bibfield  {journal} {\bibinfo  {journal} {Nature}\ }\textbf {\bibinfo {volume} {406}},\ \bibinfo {pages} {378} (\bibinfo {year} {2000})}\BibitemShut {NoStop}%
\bibitem [{\citenamefont {Callaway}\ \emph {et~al.}(2000)\citenamefont {Callaway}, \citenamefont {Newman}, \citenamefont {Strogatz},\ and\ \citenamefont {Watts}}]{Callaway2000}%
  \BibitemOpen
  \bibfield  {author} {\bibinfo {author} {\bibfnamefont {D.~S.}\ \bibnamefont {Callaway}}, \bibinfo {author} {\bibfnamefont {M.~E.~J.}\ \bibnamefont {Newman}}, \bibinfo {author} {\bibfnamefont {S.~H.}\ \bibnamefont {Strogatz}},\ and\ \bibinfo {author} {\bibfnamefont {D.~J.}\ \bibnamefont {Watts}},\ }\bibfield  {title} {\bibinfo {title} {Network robustness and fragility: Percolation on random graphs},\ }\href {https://doi.org/10.1103/PhysRevLett.85.5468} {\bibfield  {journal} {\bibinfo  {journal} {Phys. Rev. Lett.}\ }\textbf {\bibinfo {volume} {85}},\ \bibinfo {pages} {5468} (\bibinfo {year} {2000})}\BibitemShut {NoStop}%
\bibitem [{\citenamefont {Jeong}\ \emph {et~al.}(2001)\citenamefont {Jeong}, \citenamefont {Mason}, \citenamefont {Barab{\'a}si},\ and\ \citenamefont {Oltvai}}]{jeong2001lethality}%
  \BibitemOpen
  \bibfield  {author} {\bibinfo {author} {\bibfnamefont {H.}~\bibnamefont {Jeong}}, \bibinfo {author} {\bibfnamefont {S.~P.}\ \bibnamefont {Mason}}, \bibinfo {author} {\bibfnamefont {A.-L.}\ \bibnamefont {Barab{\'a}si}},\ and\ \bibinfo {author} {\bibfnamefont {Z.~N.}\ \bibnamefont {Oltvai}},\ }\bibfield  {title} {\bibinfo {title} {Lethality and centrality in protein networks},\ }\href@noop {} {\bibfield  {journal} {\bibinfo  {journal} {Nature}\ }\textbf {\bibinfo {volume} {411}},\ \bibinfo {pages} {41} (\bibinfo {year} {2001})}\BibitemShut {NoStop}%
\bibitem [{\citenamefont {De~Domenico}\ \emph {et~al.}(2014)\citenamefont {De~Domenico}, \citenamefont {Sol{\'e}-Ribalta}, \citenamefont {G{\'o}mez},\ and\ \citenamefont {Arenas}}]{de2014navigability}%
  \BibitemOpen
  \bibfield  {author} {\bibinfo {author} {\bibfnamefont {M.}~\bibnamefont {De~Domenico}}, \bibinfo {author} {\bibfnamefont {A.}~\bibnamefont {Sol{\'e}-Ribalta}}, \bibinfo {author} {\bibfnamefont {S.}~\bibnamefont {G{\'o}mez}},\ and\ \bibinfo {author} {\bibfnamefont {A.}~\bibnamefont {Arenas}},\ }\bibfield  {title} {\bibinfo {title} {Navigability of interconnected networks under random failures},\ }\href@noop {} {\bibfield  {journal} {\bibinfo  {journal} {Proceedings of the National Academy of Sciences}\ }\textbf {\bibinfo {volume} {111}},\ \bibinfo {pages} {8351} (\bibinfo {year} {2014})}\BibitemShut {NoStop}%
\bibitem [{\citenamefont {Boccaletti}\ \emph {et~al.}(2006)\citenamefont {Boccaletti}, \citenamefont {Latora}, \citenamefont {Moreno}, \citenamefont {Chavez},\ and\ \citenamefont {Hwang}}]{boccaletti2006complex}%
  \BibitemOpen
  \bibfield  {author} {\bibinfo {author} {\bibfnamefont {S.}~\bibnamefont {Boccaletti}}, \bibinfo {author} {\bibfnamefont {V.}~\bibnamefont {Latora}}, \bibinfo {author} {\bibfnamefont {Y.}~\bibnamefont {Moreno}}, \bibinfo {author} {\bibfnamefont {M.}~\bibnamefont {Chavez}},\ and\ \bibinfo {author} {\bibfnamefont {D.-U.}\ \bibnamefont {Hwang}},\ }\bibfield  {title} {\bibinfo {title} {Complex networks: Structure and dynamics},\ }\href@noop {} {\bibfield  {journal} {\bibinfo  {journal} {Physics reports}\ }\textbf {\bibinfo {volume} {424}},\ \bibinfo {pages} {175} (\bibinfo {year} {2006})}\BibitemShut {NoStop}%
\bibitem [{\citenamefont {Goltsev}\ \emph {et~al.}(2012)\citenamefont {Goltsev}, \citenamefont {Dorogovtsev}, \citenamefont {Oliveira},\ and\ \citenamefont {Mendes}}]{Goltsev2012}%
  \BibitemOpen
  \bibfield  {author} {\bibinfo {author} {\bibfnamefont {A.~V.}\ \bibnamefont {Goltsev}}, \bibinfo {author} {\bibfnamefont {S.~N.}\ \bibnamefont {Dorogovtsev}}, \bibinfo {author} {\bibfnamefont {J.~G.}\ \bibnamefont {Oliveira}},\ and\ \bibinfo {author} {\bibfnamefont {J.~F.~F.}\ \bibnamefont {Mendes}},\ }\bibfield  {title} {\bibinfo {title} {Localization and spreading of diseases in complex networks},\ }\href {https://doi.org/10.1103/PhysRevLett.109.128702} {\bibfield  {journal} {\bibinfo  {journal} {Phys. Rev. Lett.}\ }\textbf {\bibinfo {volume} {109}},\ \bibinfo {pages} {128702} (\bibinfo {year} {2012})}\BibitemShut {NoStop}%
\bibitem [{\citenamefont {Gao}\ \emph {et~al.}(2016)\citenamefont {Gao}, \citenamefont {Barzel},\ and\ \citenamefont {Barabási}}]{Gao2016}%
  \BibitemOpen
  \bibfield  {author} {\bibinfo {author} {\bibfnamefont {J.}~\bibnamefont {Gao}}, \bibinfo {author} {\bibfnamefont {B.}~\bibnamefont {Barzel}},\ and\ \bibinfo {author} {\bibfnamefont {A.-L.}\ \bibnamefont {Barabási}},\ }\bibfield  {title} {\bibinfo {title} {Universal resilience patterns in complex networks},\ }\href {https://doi.org/10.1038/nature16948} {\bibfield  {journal} {\bibinfo  {journal} {Nature}\ }\textbf {\bibinfo {volume} {530}},\ \bibinfo {pages} {307} (\bibinfo {year} {2016})}\BibitemShut {NoStop}%
\bibitem [{\citenamefont {Crua~Asensio}\ \emph {et~al.}(2017)\citenamefont {Crua~Asensio}, \citenamefont {Muñoz~Giner}, \citenamefont {de~Groot},\ and\ \citenamefont {Torrent~Burgas}}]{CruaAsensio2017}%
  \BibitemOpen
  \bibfield  {author} {\bibinfo {author} {\bibfnamefont {N.}~\bibnamefont {Crua~Asensio}}, \bibinfo {author} {\bibfnamefont {E.}~\bibnamefont {Muñoz~Giner}}, \bibinfo {author} {\bibfnamefont {N.~S.}\ \bibnamefont {de~Groot}},\ and\ \bibinfo {author} {\bibfnamefont {M.}~\bibnamefont {Torrent~Burgas}},\ }\bibfield  {title} {\bibinfo {title} {Centrality in the host–pathogen interactome is associated with pathogen fitness during infection},\ }\href {https://doi.org/10.1038/ncomms14092} {\bibfield  {journal} {\bibinfo  {journal} {Nature Communications}\ }\textbf {\bibinfo {volume} {8}},\ \bibinfo {pages} {14092} (\bibinfo {year} {2017})}\BibitemShut {NoStop}%
\bibitem [{\citenamefont {Farooq}\ \emph {et~al.}(2019)\citenamefont {Farooq}, \citenamefont {Chen}, \citenamefont {Georgiou}, \citenamefont {Tannenbaum},\ and\ \citenamefont {Lenglet}}]{Farooq2019}%
  \BibitemOpen
  \bibfield  {author} {\bibinfo {author} {\bibfnamefont {H.}~\bibnamefont {Farooq}}, \bibinfo {author} {\bibfnamefont {Y.}~\bibnamefont {Chen}}, \bibinfo {author} {\bibfnamefont {T.~T.}\ \bibnamefont {Georgiou}}, \bibinfo {author} {\bibfnamefont {A.}~\bibnamefont {Tannenbaum}},\ and\ \bibinfo {author} {\bibfnamefont {C.}~\bibnamefont {Lenglet}},\ }\bibfield  {title} {\bibinfo {title} {Network curvature as a hallmark of brain structural connectivity},\ }\href {https://doi.org/10.1038/s41467-019-12915-x} {\bibfield  {journal} {\bibinfo  {journal} {Nature Communications}\ }\textbf {\bibinfo {volume} {10}},\ \bibinfo {pages} {4937} (\bibinfo {year} {2019})}\BibitemShut {NoStop}%
\bibitem [{\citenamefont {Guilbeault}\ and\ \citenamefont {Centola}(2021)}]{Guilbeault2021}%
  \BibitemOpen
  \bibfield  {author} {\bibinfo {author} {\bibfnamefont {D.}~\bibnamefont {Guilbeault}}\ and\ \bibinfo {author} {\bibfnamefont {D.}~\bibnamefont {Centola}},\ }\bibfield  {title} {\bibinfo {title} {Topological measures for identifying and predicting the spread of complex contagions},\ }\href {https://doi.org/10.1038/s41467-021-24704-6} {\bibfield  {journal} {\bibinfo  {journal} {Nature Communications}\ }\textbf {\bibinfo {volume} {12}},\ \bibinfo {pages} {4430} (\bibinfo {year} {2021})}\BibitemShut {NoStop}%
\bibitem [{\citenamefont {Page}\ \emph {et~al.}(1999)\citenamefont {Page}, \citenamefont {Brin}, \citenamefont {Motwani},\ and\ \citenamefont {Winograd}}]{page1999pagerank}%
  \BibitemOpen
  \bibfield  {author} {\bibinfo {author} {\bibfnamefont {L.}~\bibnamefont {Page}}, \bibinfo {author} {\bibfnamefont {S.}~\bibnamefont {Brin}}, \bibinfo {author} {\bibfnamefont {R.}~\bibnamefont {Motwani}},\ and\ \bibinfo {author} {\bibfnamefont {T.}~\bibnamefont {Winograd}},\ }\href@noop {} {\emph {\bibinfo {title} {The PageRank citation ranking: Bringing order to the web.}}},\ \bibinfo {type} {Tech. Rep.}\ (\bibinfo  {institution} {Stanford InfoLab},\ \bibinfo {year} {1999})\BibitemShut {NoStop}%
\bibitem [{\citenamefont {Kitsak}\ \emph {et~al.}(2010)\citenamefont {Kitsak}, \citenamefont {Gallos}, \citenamefont {Havlin}, \citenamefont {Liljeros}, \citenamefont {Muchnik}, \citenamefont {Stanley},\ and\ \citenamefont {Makse}}]{Kitsak2010}%
  \BibitemOpen
  \bibfield  {author} {\bibinfo {author} {\bibfnamefont {M.}~\bibnamefont {Kitsak}}, \bibinfo {author} {\bibfnamefont {L.~K.}\ \bibnamefont {Gallos}}, \bibinfo {author} {\bibfnamefont {S.}~\bibnamefont {Havlin}}, \bibinfo {author} {\bibfnamefont {F.}~\bibnamefont {Liljeros}}, \bibinfo {author} {\bibfnamefont {L.}~\bibnamefont {Muchnik}}, \bibinfo {author} {\bibfnamefont {H.~E.}\ \bibnamefont {Stanley}},\ and\ \bibinfo {author} {\bibfnamefont {H.~A.}\ \bibnamefont {Makse}},\ }\bibfield  {title} {\bibinfo {title} {Identification of influential spreaders in complex networks},\ }\href {https://doi.org/10.1038/nphys1746} {\bibfield  {journal} {\bibinfo  {journal} {Nature Physics}\ }\textbf {\bibinfo {volume} {6}},\ \bibinfo {pages} {888} (\bibinfo {year} {2010})}\BibitemShut {NoStop}%
\bibitem [{\citenamefont {Salathé}\ \emph {et~al.}(2010)\citenamefont {Salathé}, \citenamefont {Kazandjieva}, \citenamefont {Lee}, \citenamefont {Levis}, \citenamefont {Feldman},\ and\ \citenamefont {Jones}}]{Salathe2010}%
  \BibitemOpen
  \bibfield  {author} {\bibinfo {author} {\bibfnamefont {M.}~\bibnamefont {Salathé}}, \bibinfo {author} {\bibfnamefont {M.}~\bibnamefont {Kazandjieva}}, \bibinfo {author} {\bibfnamefont {J.~W.}\ \bibnamefont {Lee}}, \bibinfo {author} {\bibfnamefont {P.}~\bibnamefont {Levis}}, \bibinfo {author} {\bibfnamefont {M.~W.}\ \bibnamefont {Feldman}},\ and\ \bibinfo {author} {\bibfnamefont {J.~H.}\ \bibnamefont {Jones}},\ }\bibfield  {title} {\bibinfo {title} {A high-resolution human contact network for infectious disease transmission},\ }\href {https://doi.org/10.1073/pnas.1009094108} {\bibfield  {journal} {\bibinfo  {journal} {Proceedings of the National Academy of Sciences}\ }\textbf {\bibinfo {volume} {107}},\ \bibinfo {pages} {22020} (\bibinfo {year} {2010})},\ \Eprint {https://arxiv.org/abs/https://www.pnas.org/doi/pdf/10.1073/pnas.1009094108} {https://www.pnas.org/doi/pdf/10.1073/pnas.1009094108} \BibitemShut {NoStop}%
\bibitem [{\citenamefont {Wang}\ \emph {et~al.}(2016)\citenamefont {Wang}, \citenamefont {Bauch}, \citenamefont {Bhattacharyya}, \citenamefont {d'Onofrio}, \citenamefont {Manfredi}, \citenamefont {Perc}, \citenamefont {Perra}, \citenamefont {Salathé},\ and\ \citenamefont {Zhao}}]{Wang2016}%
  \BibitemOpen
  \bibfield  {author} {\bibinfo {author} {\bibfnamefont {Z.}~\bibnamefont {Wang}}, \bibinfo {author} {\bibfnamefont {C.~T.}\ \bibnamefont {Bauch}}, \bibinfo {author} {\bibfnamefont {S.}~\bibnamefont {Bhattacharyya}}, \bibinfo {author} {\bibfnamefont {A.}~\bibnamefont {d'Onofrio}}, \bibinfo {author} {\bibfnamefont {P.}~\bibnamefont {Manfredi}}, \bibinfo {author} {\bibfnamefont {M.}~\bibnamefont {Perc}}, \bibinfo {author} {\bibfnamefont {N.}~\bibnamefont {Perra}}, \bibinfo {author} {\bibfnamefont {M.}~\bibnamefont {Salathé}},\ and\ \bibinfo {author} {\bibfnamefont {D.}~\bibnamefont {Zhao}},\ }\bibfield  {title} {\bibinfo {title} {Statistical physics of vaccination},\ }\href {https://doi.org/https://doi.org/10.1016/j.physrep.2016.10.006} {\bibfield  {journal} {\bibinfo  {journal} {Physics Reports}\ }\textbf {\bibinfo {volume} {664}},\ \bibinfo {pages} {1} (\bibinfo {year} {2016})},\ \bibinfo {note} {statistical physics of vaccination}\BibitemShut {NoStop}%
\bibitem [{\citenamefont {Pung}\ \emph {et~al.}(2022)\citenamefont {Pung}, \citenamefont {Firth}, \citenamefont {Spurgin}, \citenamefont {{Singapore CruiseSafe working group}},\ and\ \citenamefont {{CMMID COVID-19 working group}}}]{Pung2022}%
  \BibitemOpen
  \bibfield  {author} {\bibinfo {author} {\bibfnamefont {R.}~\bibnamefont {Pung}}, \bibinfo {author} {\bibfnamefont {J.~A.}\ \bibnamefont {Firth}}, \bibinfo {author} {\bibnamefont {Spurgin}}, \bibinfo {author} {\bibnamefont {{Singapore CruiseSafe working group}}},\ and\ \bibinfo {author} {\bibnamefont {{CMMID COVID-19 working group}}},\ }\bibfield  {title} {\bibinfo {title} {Using high-resolution contact networks to evaluate {SARS-CoV-2} transmission and control in large-scale multi-day events},\ }\href {https://doi.org/10.1038/s41467-022-29522-y} {\bibfield  {journal} {\bibinfo  {journal} {Nature Communications}\ }\textbf {\bibinfo {volume} {13}},\ \bibinfo {pages} {1956} (\bibinfo {year} {2022})}\BibitemShut {NoStop}%
\bibitem [{\citenamefont {Guimerà}\ \emph {et~al.}(2005)\citenamefont {Guimerà}, \citenamefont {Mossa}, \citenamefont {Turtschi},\ and\ \citenamefont {Amaral}}]{Guimera2005}%
  \BibitemOpen
  \bibfield  {author} {\bibinfo {author} {\bibfnamefont {R.}~\bibnamefont {Guimerà}}, \bibinfo {author} {\bibfnamefont {S.}~\bibnamefont {Mossa}}, \bibinfo {author} {\bibfnamefont {A.}~\bibnamefont {Turtschi}},\ and\ \bibinfo {author} {\bibfnamefont {L.~A.~N.}\ \bibnamefont {Amaral}},\ }\bibfield  {title} {\bibinfo {title} {The worldwide air transportation network: Anomalous centrality, community structure, and cities' global roles},\ }\href {https://doi.org/10.1073/pnas.0407994102} {\bibfield  {journal} {\bibinfo  {journal} {Proceedings of the National Academy of Sciences}\ }\textbf {\bibinfo {volume} {102}},\ \bibinfo {pages} {7794} (\bibinfo {year} {2005})},\ \Eprint {https://arxiv.org/abs/https://www.pnas.org/doi/pdf/10.1073/pnas.0407994102} {https://www.pnas.org/doi/pdf/10.1073/pnas.0407994102} \BibitemShut {NoStop}%
\bibitem [{\citenamefont {Wu}\ \emph {et~al.}(2006)\citenamefont {Wu}, \citenamefont {Braunstein}, \citenamefont {Havlin},\ and\ \citenamefont {Stanley}}]{Wu2006}%
  \BibitemOpen
  \bibfield  {author} {\bibinfo {author} {\bibfnamefont {Z.}~\bibnamefont {Wu}}, \bibinfo {author} {\bibfnamefont {L.~A.}\ \bibnamefont {Braunstein}}, \bibinfo {author} {\bibfnamefont {S.}~\bibnamefont {Havlin}},\ and\ \bibinfo {author} {\bibfnamefont {H.~E.}\ \bibnamefont {Stanley}},\ }\bibfield  {title} {\bibinfo {title} {Transport in weighted networks: Partition into superhighways and roads},\ }\href {https://doi.org/10.1103/PhysRevLett.96.148702} {\bibfield  {journal} {\bibinfo  {journal} {Phys. Rev. Lett.}\ }\textbf {\bibinfo {volume} {96}},\ \bibinfo {pages} {148702} (\bibinfo {year} {2006})}\BibitemShut {NoStop}%
\bibitem [{\citenamefont {Brown}\ \emph {et~al.}(2006)\citenamefont {Brown}, \citenamefont {Carlyle}, \citenamefont {Salmer{\'o}n},\ and\ \citenamefont {Wood}}]{brown2006defending}%
  \BibitemOpen
  \bibfield  {author} {\bibinfo {author} {\bibfnamefont {G.}~\bibnamefont {Brown}}, \bibinfo {author} {\bibfnamefont {M.}~\bibnamefont {Carlyle}}, \bibinfo {author} {\bibfnamefont {J.}~\bibnamefont {Salmer{\'o}n}},\ and\ \bibinfo {author} {\bibfnamefont {K.}~\bibnamefont {Wood}},\ }\bibfield  {title} {\bibinfo {title} {Defending critical infrastructure},\ }\href@noop {} {\bibfield  {journal} {\bibinfo  {journal} {Interfaces}\ }\textbf {\bibinfo {volume} {36}},\ \bibinfo {pages} {530} (\bibinfo {year} {2006})}\BibitemShut {NoStop}%
\bibitem [{\citenamefont {Carvalho}\ \emph {et~al.}(2009)\citenamefont {Carvalho}, \citenamefont {Buzna}, \citenamefont {Bono}, \citenamefont {Guti\'errez}, \citenamefont {Just},\ and\ \citenamefont {Arrowsmith}}]{Carvalho2009}%
  \BibitemOpen
  \bibfield  {author} {\bibinfo {author} {\bibfnamefont {R.}~\bibnamefont {Carvalho}}, \bibinfo {author} {\bibfnamefont {L.}~\bibnamefont {Buzna}}, \bibinfo {author} {\bibfnamefont {F.}~\bibnamefont {Bono}}, \bibinfo {author} {\bibfnamefont {E.}~\bibnamefont {Guti\'errez}}, \bibinfo {author} {\bibfnamefont {W.}~\bibnamefont {Just}},\ and\ \bibinfo {author} {\bibfnamefont {D.}~\bibnamefont {Arrowsmith}},\ }\bibfield  {title} {\bibinfo {title} {Robustness of trans-european gas networks},\ }\href {https://doi.org/10.1103/PhysRevE.80.016106} {\bibfield  {journal} {\bibinfo  {journal} {Phys. Rev. E}\ }\textbf {\bibinfo {volume} {80}},\ \bibinfo {pages} {016106} (\bibinfo {year} {2009})}\BibitemShut {NoStop}%
\bibitem [{\citenamefont {Duan}\ and\ \citenamefont {Lu}(2014)}]{Duan2014}%
  \BibitemOpen
  \bibfield  {author} {\bibinfo {author} {\bibfnamefont {Y.}~\bibnamefont {Duan}}\ and\ \bibinfo {author} {\bibfnamefont {F.}~\bibnamefont {Lu}},\ }\bibfield  {title} {\bibinfo {title} {Robustness of city road networks at different granularities},\ }\href {https://doi.org/https://doi.org/10.1016/j.physa.2014.05.073} {\bibfield  {journal} {\bibinfo  {journal} {Physica A: Statistical Mechanics and its Applications}\ }\textbf {\bibinfo {volume} {411}},\ \bibinfo {pages} {21} (\bibinfo {year} {2014})}\BibitemShut {NoStop}%
\bibitem [{\citenamefont {Bonacich}(1972)}]{bonacich1972factoring}%
  \BibitemOpen
  \bibfield  {author} {\bibinfo {author} {\bibfnamefont {P.}~\bibnamefont {Bonacich}},\ }\bibfield  {title} {\bibinfo {title} {Factoring and weighting approaches to status scores and clique identification},\ }\href@noop {} {\bibfield  {journal} {\bibinfo  {journal} {Journal of mathematical sociology}\ }\textbf {\bibinfo {volume} {2}},\ \bibinfo {pages} {113} (\bibinfo {year} {1972})}\BibitemShut {NoStop}%
\bibitem [{\citenamefont {Katz}(1953)}]{katz1953new}%
  \BibitemOpen
  \bibfield  {author} {\bibinfo {author} {\bibfnamefont {L.}~\bibnamefont {Katz}},\ }\bibfield  {title} {\bibinfo {title} {A new status index derived from sociometric analysis},\ }\href@noop {} {\bibfield  {journal} {\bibinfo  {journal} {Psychometrika}\ }\textbf {\bibinfo {volume} {18}},\ \bibinfo {pages} {39} (\bibinfo {year} {1953})}\BibitemShut {NoStop}%
\bibitem [{\citenamefont {Freeman}(1977)}]{freeman1977set}%
  \BibitemOpen
  \bibfield  {author} {\bibinfo {author} {\bibfnamefont {L.~C.}\ \bibnamefont {Freeman}},\ }\bibfield  {title} {\bibinfo {title} {A set of measures of centrality based on betweenness},\ }\href@noop {} {\bibfield  {journal} {\bibinfo  {journal} {Sociometry}\ ,\ \bibinfo {pages} {35}} (\bibinfo {year} {1977})}\BibitemShut {NoStop}%
\bibitem [{\citenamefont {Brandes}\ and\ \citenamefont {Fleischer}(2005)}]{brandes2005centrality}%
  \BibitemOpen
  \bibfield  {author} {\bibinfo {author} {\bibfnamefont {U.}~\bibnamefont {Brandes}}\ and\ \bibinfo {author} {\bibfnamefont {D.}~\bibnamefont {Fleischer}},\ }\bibfield  {title} {\bibinfo {title} {Centrality measures based on current flow.},\ }in\ \href@noop {} {\emph {\bibinfo {booktitle} {Stacs}}},\ Vol.\ \bibinfo {volume} {3404}\ (\bibinfo {organization} {Springer},\ \bibinfo {year} {2005})\ pp.\ \bibinfo {pages} {533--544}\BibitemShut {NoStop}%
\bibitem [{\citenamefont {Ghavasieh}\ \emph {et~al.}(2021)\citenamefont {Ghavasieh}, \citenamefont {Stella}, \citenamefont {Biamonte},\ and\ \citenamefont {De~Domenico}}]{ghavasieh2021unraveling}%
  \BibitemOpen
  \bibfield  {author} {\bibinfo {author} {\bibfnamefont {A.}~\bibnamefont {Ghavasieh}}, \bibinfo {author} {\bibfnamefont {M.}~\bibnamefont {Stella}}, \bibinfo {author} {\bibfnamefont {J.}~\bibnamefont {Biamonte}},\ and\ \bibinfo {author} {\bibfnamefont {M.}~\bibnamefont {De~Domenico}},\ }\bibfield  {title} {\bibinfo {title} {Unraveling the effects of multiscale network entanglement on empirical systems},\ }\href@noop {} {\bibfield  {journal} {\bibinfo  {journal} {Communications Physics}\ }\textbf {\bibinfo {volume} {4}},\ \bibinfo {pages} {129} (\bibinfo {year} {2021})}\BibitemShut {NoStop}%
\bibitem [{\citenamefont {de~Arruda}\ \emph {et~al.}(2014)\citenamefont {de~Arruda}, \citenamefont {Barbieri}, \citenamefont {Rodr\'{\i}guez}, \citenamefont {Rodrigues}, \citenamefont {Moreno},\ and\ \citenamefont {Costa}}]{PhysRevE.90.032812}%
  \BibitemOpen
  \bibfield  {author} {\bibinfo {author} {\bibfnamefont {G.~F.}\ \bibnamefont {de~Arruda}}, \bibinfo {author} {\bibfnamefont {A.~L.}\ \bibnamefont {Barbieri}}, \bibinfo {author} {\bibfnamefont {P.~M.}\ \bibnamefont {Rodr\'{\i}guez}}, \bibinfo {author} {\bibfnamefont {F.~A.}\ \bibnamefont {Rodrigues}}, \bibinfo {author} {\bibfnamefont {Y.}~\bibnamefont {Moreno}},\ and\ \bibinfo {author} {\bibfnamefont {L.~d.~F.}\ \bibnamefont {Costa}},\ }\bibfield  {title} {\bibinfo {title} {Role of centrality for the identification of influential spreaders in complex networks},\ }\href {https://doi.org/10.1103/PhysRevE.90.032812} {\bibfield  {journal} {\bibinfo  {journal} {Phys. Rev. E}\ }\textbf {\bibinfo {volume} {90}},\ \bibinfo {pages} {032812} (\bibinfo {year} {2014})}\BibitemShut {NoStop}%
\bibitem [{\citenamefont {Sol\'e}\ \emph {et~al.}(2008)\citenamefont {Sol\'e}, \citenamefont {Rosas-Casals}, \citenamefont {Corominas-Murtra},\ and\ \citenamefont {Valverde}}]{Sole2008}%
  \BibitemOpen
  \bibfield  {author} {\bibinfo {author} {\bibfnamefont {R.~V.}\ \bibnamefont {Sol\'e}}, \bibinfo {author} {\bibfnamefont {M.}~\bibnamefont {Rosas-Casals}}, \bibinfo {author} {\bibfnamefont {B.}~\bibnamefont {Corominas-Murtra}},\ and\ \bibinfo {author} {\bibfnamefont {S.}~\bibnamefont {Valverde}},\ }\bibfield  {title} {\bibinfo {title} {Robustness of the european power grids under intentional attack},\ }\href {https://doi.org/10.1103/PhysRevE.77.026102} {\bibfield  {journal} {\bibinfo  {journal} {Phys. Rev. E}\ }\textbf {\bibinfo {volume} {77}},\ \bibinfo {pages} {026102} (\bibinfo {year} {2008})}\BibitemShut {NoStop}%
\bibitem [{\citenamefont {Doyle}\ \emph {et~al.}(2005)\citenamefont {Doyle}, \citenamefont {Alderson}, \citenamefont {Li}, \citenamefont {Low}, \citenamefont {Roughan}, \citenamefont {Shalunov}, \citenamefont {Tanaka},\ and\ \citenamefont {Willinger}}]{doyle2005robust}%
  \BibitemOpen
  \bibfield  {author} {\bibinfo {author} {\bibfnamefont {J.~C.}\ \bibnamefont {Doyle}}, \bibinfo {author} {\bibfnamefont {D.~L.}\ \bibnamefont {Alderson}}, \bibinfo {author} {\bibfnamefont {L.}~\bibnamefont {Li}}, \bibinfo {author} {\bibfnamefont {S.}~\bibnamefont {Low}}, \bibinfo {author} {\bibfnamefont {M.}~\bibnamefont {Roughan}}, \bibinfo {author} {\bibfnamefont {S.}~\bibnamefont {Shalunov}}, \bibinfo {author} {\bibfnamefont {R.}~\bibnamefont {Tanaka}},\ and\ \bibinfo {author} {\bibfnamefont {W.}~\bibnamefont {Willinger}},\ }\bibfield  {title} {\bibinfo {title} {The “robust yet fragile” nature of the internet},\ }\href@noop {} {\bibfield  {journal} {\bibinfo  {journal} {Proceedings of the National Academy of Sciences}\ }\textbf {\bibinfo {volume} {102}},\ \bibinfo {pages} {14497} (\bibinfo {year} {2005})}\BibitemShut {NoStop}%
\bibitem [{\citenamefont {Rinaldi}\ \emph {et~al.}(2001)\citenamefont {Rinaldi}, \citenamefont {Peerenboom},\ and\ \citenamefont {Kelly}}]{rinaldi2001identifying}%
  \BibitemOpen
  \bibfield  {author} {\bibinfo {author} {\bibfnamefont {S.~M.}\ \bibnamefont {Rinaldi}}, \bibinfo {author} {\bibfnamefont {J.~P.}\ \bibnamefont {Peerenboom}},\ and\ \bibinfo {author} {\bibfnamefont {T.~K.}\ \bibnamefont {Kelly}},\ }\bibfield  {title} {\bibinfo {title} {Identifying, understanding, and analyzing critical infrastructure interdependencies},\ }\href@noop {} {\bibfield  {journal} {\bibinfo  {journal} {IEEE control systems magazine}\ }\textbf {\bibinfo {volume} {21}},\ \bibinfo {pages} {11} (\bibinfo {year} {2001})}\BibitemShut {NoStop}%
\bibitem [{\citenamefont {Cohen}\ \emph {et~al.}(2000)\citenamefont {Cohen}, \citenamefont {Erez}, \citenamefont {ben Avraham},\ and\ \citenamefont {Havlin}}]{Cohen20001}%
  \BibitemOpen
  \bibfield  {author} {\bibinfo {author} {\bibfnamefont {R.}~\bibnamefont {Cohen}}, \bibinfo {author} {\bibfnamefont {K.}~\bibnamefont {Erez}}, \bibinfo {author} {\bibfnamefont {D.}~\bibnamefont {ben Avraham}},\ and\ \bibinfo {author} {\bibfnamefont {S.}~\bibnamefont {Havlin}},\ }\bibfield  {title} {\bibinfo {title} {Resilience of the internet to random breakdowns},\ }\href {https://doi.org/10.1103/PhysRevLett.85.4626} {\bibfield  {journal} {\bibinfo  {journal} {Phys. Rev. Lett.}\ }\textbf {\bibinfo {volume} {85}},\ \bibinfo {pages} {4626} (\bibinfo {year} {2000})}\BibitemShut {NoStop}%
\bibitem [{\citenamefont {Schäfer}\ \emph {et~al.}(2018)\citenamefont {Schäfer}, \citenamefont {Witthaut}, \citenamefont {Timme},\ and\ \citenamefont {Latora}}]{Schafer2018}%
  \BibitemOpen
  \bibfield  {author} {\bibinfo {author} {\bibfnamefont {B.}~\bibnamefont {Schäfer}}, \bibinfo {author} {\bibfnamefont {D.}~\bibnamefont {Witthaut}}, \bibinfo {author} {\bibfnamefont {M.}~\bibnamefont {Timme}},\ and\ \bibinfo {author} {\bibfnamefont {V.}~\bibnamefont {Latora}},\ }\bibfield  {title} {\bibinfo {title} {Dynamically induced cascading failures in power grids},\ }\href {https://doi.org/10.1038/s41467-018-04287-5} {\bibfield  {journal} {\bibinfo  {journal} {Nature Communications}\ }\textbf {\bibinfo {volume} {9}},\ \bibinfo {pages} {1975} (\bibinfo {year} {2018})}\BibitemShut {NoStop}%
\bibitem [{\citenamefont {Colizza}\ \emph {et~al.}(2006)\citenamefont {Colizza}, \citenamefont {Flammini}, \citenamefont {Serrano},\ and\ \citenamefont {Vespignani}}]{colizza2006detecting}%
  \BibitemOpen
  \bibfield  {author} {\bibinfo {author} {\bibfnamefont {V.}~\bibnamefont {Colizza}}, \bibinfo {author} {\bibfnamefont {A.}~\bibnamefont {Flammini}}, \bibinfo {author} {\bibfnamefont {M.~A.}\ \bibnamefont {Serrano}},\ and\ \bibinfo {author} {\bibfnamefont {A.}~\bibnamefont {Vespignani}},\ }\bibfield  {title} {\bibinfo {title} {Detecting rich-club ordering in complex networks},\ }\href@noop {} {\bibfield  {journal} {\bibinfo  {journal} {Nature physics}\ }\textbf {\bibinfo {volume} {2}},\ \bibinfo {pages} {110} (\bibinfo {year} {2006})}\BibitemShut {NoStop}%
\bibitem [{\citenamefont {McAuley}\ \emph {et~al.}(2007)\citenamefont {McAuley}, \citenamefont {da~Fontoura~Costa},\ and\ \citenamefont {Caetano}}]{mcauley2007rich}%
  \BibitemOpen
  \bibfield  {author} {\bibinfo {author} {\bibfnamefont {J.~J.}\ \bibnamefont {McAuley}}, \bibinfo {author} {\bibfnamefont {L.}~\bibnamefont {da~Fontoura~Costa}},\ and\ \bibinfo {author} {\bibfnamefont {T.~S.}\ \bibnamefont {Caetano}},\ }\bibfield  {title} {\bibinfo {title} {Rich-club phenomenon across complex network hierarchies},\ }\href@noop {} {\bibfield  {journal} {\bibinfo  {journal} {Applied Physics Letters}\ }\textbf {\bibinfo {volume} {91}} (\bibinfo {year} {2007})}\BibitemShut {NoStop}%
\bibitem [{\citenamefont {Van Den~Heuvel}\ and\ \citenamefont {Sporns}(2011)}]{van2011rich}%
  \BibitemOpen
  \bibfield  {author} {\bibinfo {author} {\bibfnamefont {M.~P.}\ \bibnamefont {Van Den~Heuvel}}\ and\ \bibinfo {author} {\bibfnamefont {O.}~\bibnamefont {Sporns}},\ }\bibfield  {title} {\bibinfo {title} {Rich-club organization of the human connectome},\ }\href@noop {} {\bibfield  {journal} {\bibinfo  {journal} {Journal of Neuroscience}\ }\textbf {\bibinfo {volume} {31}},\ \bibinfo {pages} {15775} (\bibinfo {year} {2011})}\BibitemShut {NoStop}%
\bibitem [{\citenamefont {Alstott}\ \emph {et~al.}(2014)\citenamefont {Alstott}, \citenamefont {Panzarasa}, \citenamefont {Rubinov}, \citenamefont {Bullmore},\ and\ \citenamefont {V{\'e}rtes}}]{Alstott2014}%
  \BibitemOpen
  \bibfield  {author} {\bibinfo {author} {\bibfnamefont {J.}~\bibnamefont {Alstott}}, \bibinfo {author} {\bibfnamefont {P.}~\bibnamefont {Panzarasa}}, \bibinfo {author} {\bibfnamefont {M.}~\bibnamefont {Rubinov}}, \bibinfo {author} {\bibfnamefont {E.~T.}\ \bibnamefont {Bullmore}},\ and\ \bibinfo {author} {\bibfnamefont {P.~E.}\ \bibnamefont {V{\'e}rtes}},\ }\bibfield  {title} {\bibinfo {title} {A unifying framework for measuring weighted rich clubs},\ }\href {https://doi.org/10.1038/srep07258} {\bibfield  {journal} {\bibinfo  {journal} {Scientific Reports}\ }\textbf {\bibinfo {volume} {4}},\ \bibinfo {pages} {7258} (\bibinfo {year} {2014})}\BibitemShut {NoStop}%
\bibitem [{\citenamefont {Trajanovski}\ \emph {et~al.}(2013)\citenamefont {Trajanovski}, \citenamefont {Martín-Hernández}, \citenamefont {Winterbach},\ and\ \citenamefont {Van~Mieghem}}]{Trajanovski2013}%
  \BibitemOpen
  \bibfield  {author} {\bibinfo {author} {\bibfnamefont {S.}~\bibnamefont {Trajanovski}}, \bibinfo {author} {\bibfnamefont {J.}~\bibnamefont {Martín-Hernández}}, \bibinfo {author} {\bibfnamefont {W.}~\bibnamefont {Winterbach}},\ and\ \bibinfo {author} {\bibfnamefont {P.}~\bibnamefont {Van~Mieghem}},\ }\bibfield  {title} {\bibinfo {title} {{Robustness envelopes of networks}},\ }\href {https://doi.org/10.1093/comnet/cnt004} {\bibfield  {journal} {\bibinfo  {journal} {Journal of Complex Networks}\ }\textbf {\bibinfo {volume} {1}},\ \bibinfo {pages} {44} (\bibinfo {year} {2013})},\ \Eprint {https://arxiv.org/abs/https://academic.oup.com/comnet/article-pdf/1/1/44/1369171/cnt004.pdf} {https://academic.oup.com/comnet/article-pdf/1/1/44/1369171/cnt004.pdf} \BibitemShut {NoStop}%
\bibitem [{\citenamefont {Ventresca}\ and\ \citenamefont {Aleman}(2014)}]{Ventresca2015}%
  \BibitemOpen
  \bibfield  {author} {\bibinfo {author} {\bibfnamefont {M.}~\bibnamefont {Ventresca}}\ and\ \bibinfo {author} {\bibfnamefont {D.}~\bibnamefont {Aleman}},\ }\bibfield  {title} {\bibinfo {title} {{Network robustness versus multi-strategy sequential attack}},\ }\href {https://doi.org/10.1093/comnet/cnu010} {\bibfield  {journal} {\bibinfo  {journal} {Journal of Complex Networks}\ }\textbf {\bibinfo {volume} {3}},\ \bibinfo {pages} {126} (\bibinfo {year} {2014})},\ \Eprint {https://arxiv.org/abs/https://academic.oup.com/comnet/article-pdf/3/1/126/1323613/cnu010.pdf} {https://academic.oup.com/comnet/article-pdf/3/1/126/1323613/cnu010.pdf} \BibitemShut {NoStop}%
\bibitem [{\citenamefont {Williams}\ and\ \citenamefont {Musolesi}(2016)}]{Williams2016}%
  \BibitemOpen
  \bibfield  {author} {\bibinfo {author} {\bibfnamefont {M.~J.}\ \bibnamefont {Williams}}\ and\ \bibinfo {author} {\bibfnamefont {M.}~\bibnamefont {Musolesi}},\ }\bibfield  {title} {\bibinfo {title} {Spatio-temporal networks: reachability, centrality and robustness},\ }\href {https://doi.org/10.1098/rsos.160196} {\bibfield  {journal} {\bibinfo  {journal} {Royal Society Open Science}\ }\textbf {\bibinfo {volume} {3}},\ \bibinfo {pages} {160196} (\bibinfo {year} {2016})}\BibitemShut {NoStop}%
\bibitem [{\citenamefont {Cats}\ and\ \citenamefont {Krishnakumari}(2020)}]{Cats2020}%
  \BibitemOpen
  \bibfield  {author} {\bibinfo {author} {\bibfnamefont {O.}~\bibnamefont {Cats}}\ and\ \bibinfo {author} {\bibfnamefont {P.}~\bibnamefont {Krishnakumari}},\ }\bibfield  {title} {\bibinfo {title} {Metropolitan rail network robustness},\ }\href {https://doi.org/https://doi.org/10.1016/j.physa.2020.124317} {\bibfield  {journal} {\bibinfo  {journal} {Physica A: Statistical Mechanics and its Applications}\ }\textbf {\bibinfo {volume} {549}},\ \bibinfo {pages} {124317} (\bibinfo {year} {2020})}\BibitemShut {NoStop}%
\bibitem [{\citenamefont {Biggs}\ \emph {et~al.}(1993)\citenamefont {Biggs}, \citenamefont {Biggs},\ and\ \citenamefont {Norman}}]{biggs1993algebraic}%
  \BibitemOpen
  \bibfield  {author} {\bibinfo {author} {\bibfnamefont {N.}~\bibnamefont {Biggs}}, \bibinfo {author} {\bibfnamefont {N.~L.}\ \bibnamefont {Biggs}},\ and\ \bibinfo {author} {\bibfnamefont {B.}~\bibnamefont {Norman}},\ }\href@noop {} {\emph {\bibinfo {title} {Algebraic graph theory}}},\ \bibinfo {number} {67}\ (\bibinfo  {publisher} {Cambridge university press},\ \bibinfo {year} {1993})\BibitemShut {NoStop}%
\bibitem [{\citenamefont {Erd\"{o}s}\ and\ \citenamefont {R\'{e}nyi}(1959)}]{Erdos1959}%
  \BibitemOpen
  \bibfield  {author} {\bibinfo {author} {\bibfnamefont {P.}~\bibnamefont {Erd\"{o}s}}\ and\ \bibinfo {author} {\bibfnamefont {A.}~\bibnamefont {R\'{e}nyi}},\ }\bibfield  {title} {\bibinfo {title} {On random graphs {I}},\ }\href@noop {} {\bibfield  {journal} {\bibinfo  {journal} {Publicationes Mathematicae Debrecen}\ }\textbf {\bibinfo {volume} {6}},\ \bibinfo {pages} {290} (\bibinfo {year} {1959})}\BibitemShut {NoStop}%
\bibitem [{\citenamefont {Barab{\'a}si}\ and\ \citenamefont {Albert}(1999)}]{Barabasi1999}%
  \BibitemOpen
  \bibfield  {author} {\bibinfo {author} {\bibfnamefont {A.-L.}\ \bibnamefont {Barab{\'a}si}}\ and\ \bibinfo {author} {\bibfnamefont {R.}~\bibnamefont {Albert}},\ }\bibfield  {title} {\bibinfo {title} {Emergence of scaling in random networks},\ }\href {https://doi.org/10.1126/science.286.5439.509} {\bibfield  {journal} {\bibinfo  {journal} {Science}\ }\textbf {\bibinfo {volume} {286}},\ \bibinfo {pages} {509} (\bibinfo {year} {1999})}\BibitemShut {NoStop}%
\bibitem [{\citenamefont {Watts}\ and\ \citenamefont {Strogatz}(1998)}]{Watts1998}%
  \BibitemOpen
  \bibfield  {author} {\bibinfo {author} {\bibfnamefont {D.~J.}\ \bibnamefont {Watts}}\ and\ \bibinfo {author} {\bibfnamefont {S.~H.}\ \bibnamefont {Strogatz}},\ }\bibfield  {title} {\bibinfo {title} {Collective dynamics of ‘small-world’ networks},\ }\href@noop {} {\bibfield  {journal} {\bibinfo  {journal} {Nature}\ }\textbf {\bibinfo {volume} {393}},\ \bibinfo {pages} {440} (\bibinfo {year} {1998})}\BibitemShut {NoStop}%
\bibitem [{\citenamefont {Holland}\ \emph {et~al.}(1983)\citenamefont {Holland}, \citenamefont {Laskey},\ and\ \citenamefont {Leinhardt}}]{holland1983stochastic}%
  \BibitemOpen
  \bibfield  {author} {\bibinfo {author} {\bibfnamefont {P.~W.}\ \bibnamefont {Holland}}, \bibinfo {author} {\bibfnamefont {K.~B.}\ \bibnamefont {Laskey}},\ and\ \bibinfo {author} {\bibfnamefont {S.}~\bibnamefont {Leinhardt}},\ }\bibfield  {title} {\bibinfo {title} {Stochastic blockmodels: First steps},\ }\href@noop {} {\bibfield  {journal} {\bibinfo  {journal} {Social networks}\ }\textbf {\bibinfo {volume} {5}},\ \bibinfo {pages} {109} (\bibinfo {year} {1983})}\BibitemShut {NoStop}%
\bibitem [{\citenamefont {Gilbert}(1959)}]{gilbert1959random}%
  \BibitemOpen
  \bibfield  {author} {\bibinfo {author} {\bibfnamefont {E.~N.}\ \bibnamefont {Gilbert}},\ }\bibfield  {title} {\bibinfo {title} {Random graphs},\ }\href@noop {} {\bibfield  {journal} {\bibinfo  {journal} {The Annals of Mathematical Statistics}\ }\textbf {\bibinfo {volume} {30}},\ \bibinfo {pages} {1141} (\bibinfo {year} {1959})}\BibitemShut {NoStop}%
\bibitem [{\citenamefont {Tejedor}\ \emph {et~al.}(2017)\citenamefont {Tejedor}, \citenamefont {Longjas}, \citenamefont {Zaliapin}, \citenamefont {Ambroj},\ and\ \citenamefont {Foufoula-Georgiou}}]{Tejedor2017_AI}%
  \BibitemOpen
  \bibfield  {author} {\bibinfo {author} {\bibfnamefont {A.}~\bibnamefont {Tejedor}}, \bibinfo {author} {\bibfnamefont {A.}~\bibnamefont {Longjas}}, \bibinfo {author} {\bibfnamefont {I.}~\bibnamefont {Zaliapin}}, \bibinfo {author} {\bibfnamefont {S.}~\bibnamefont {Ambroj}},\ and\ \bibinfo {author} {\bibfnamefont {E.}~\bibnamefont {Foufoula-Georgiou}},\ }\bibfield  {title} {\bibinfo {title} {Network robustness assessed within a dual connectivity framework: joint dynamics of the active and idle networks},\ }\href {https://doi.org/10.1038/s41598-017-08714-3} {\bibfield  {journal} {\bibinfo  {journal} {Scientific Reports}\ }\textbf {\bibinfo {volume} {7}},\ \bibinfo {pages} {8567} (\bibinfo {year} {2017})}\BibitemShut {NoStop}%
\bibitem [{\citenamefont {Manke}\ \emph {et~al.}(2006)\citenamefont {Manke}, \citenamefont {Demetrius},\ and\ \citenamefont {Vingron}}]{manke2006entropic}%
  \BibitemOpen
  \bibfield  {author} {\bibinfo {author} {\bibfnamefont {T.}~\bibnamefont {Manke}}, \bibinfo {author} {\bibfnamefont {L.}~\bibnamefont {Demetrius}},\ and\ \bibinfo {author} {\bibfnamefont {M.}~\bibnamefont {Vingron}},\ }\bibfield  {title} {\bibinfo {title} {An entropic characterization of protein interaction networks and cellular robustness},\ }\href@noop {} {\bibfield  {journal} {\bibinfo  {journal} {Journal of The Royal Society Interface}\ }\textbf {\bibinfo {volume} {3}},\ \bibinfo {pages} {843} (\bibinfo {year} {2006})}\BibitemShut {NoStop}%
\bibitem [{\citenamefont {Rossi}\ and\ \citenamefont {Ahmed}(2015)}]{nr}%
  \BibitemOpen
  \bibfield  {author} {\bibinfo {author} {\bibfnamefont {R.~A.}\ \bibnamefont {Rossi}}\ and\ \bibinfo {author} {\bibfnamefont {N.~K.}\ \bibnamefont {Ahmed}},\ }\bibfield  {title} {\bibinfo {title} {The network data repository with interactive graph analytics and visualization},\ }in\ \href {https://networkrepository.com} {\emph {\bibinfo {booktitle} {AAAI}}}\ (\bibinfo {year} {2015})\BibitemShut {NoStop}%
\bibitem [{\citenamefont {Kunegis}(2013)}]{kunegis2013konect}%
  \BibitemOpen
  \bibfield  {author} {\bibinfo {author} {\bibfnamefont {J.}~\bibnamefont {Kunegis}},\ }\bibfield  {title} {\bibinfo {title} {Konect: the koblenz network collection},\ }in\ \href@noop {} {\emph {\bibinfo {booktitle} {Proceedings of the 22nd international conference on world wide web}}}\ (\bibinfo {year} {2013})\ pp.\ \bibinfo {pages} {1343--1350}\BibitemShut {NoStop}%
\bibitem [{\citenamefont {Leskovec}\ \emph {et~al.}(2007)\citenamefont {Leskovec}, \citenamefont {Kleinberg},\ and\ \citenamefont {Faloutsos}}]{leskovec2007graph}%
  \BibitemOpen
  \bibfield  {author} {\bibinfo {author} {\bibfnamefont {J.}~\bibnamefont {Leskovec}}, \bibinfo {author} {\bibfnamefont {J.}~\bibnamefont {Kleinberg}},\ and\ \bibinfo {author} {\bibfnamefont {C.}~\bibnamefont {Faloutsos}},\ }\bibfield  {title} {\bibinfo {title} {Graph evolution: Densification and shrinking diameters},\ }\href@noop {} {\bibfield  {journal} {\bibinfo  {journal} {ACM transactions on Knowledge Discovery from Data (TKDD)}\ }\textbf {\bibinfo {volume} {1}},\ \bibinfo {pages} {2} (\bibinfo {year} {2007})}\BibitemShut {NoStop}%
\bibitem [{\citenamefont {Brandes}(2001)}]{brandes2001faster}%
  \BibitemOpen
  \bibfield  {author} {\bibinfo {author} {\bibfnamefont {U.}~\bibnamefont {Brandes}},\ }\bibfield  {title} {\bibinfo {title} {A faster algorithm for betweenness centrality},\ }\href@noop {} {\bibfield  {journal} {\bibinfo  {journal} {Journal of mathematical sociology}\ }\textbf {\bibinfo {volume} {25}},\ \bibinfo {pages} {163} (\bibinfo {year} {2001})}\BibitemShut {NoStop}%
\bibitem [{\citenamefont {Morone}\ and\ \citenamefont {Makse}(2015)}]{morone2015influence}%
  \BibitemOpen
  \bibfield  {author} {\bibinfo {author} {\bibfnamefont {F.}~\bibnamefont {Morone}}\ and\ \bibinfo {author} {\bibfnamefont {H.~A.}\ \bibnamefont {Makse}},\ }\bibfield  {title} {\bibinfo {title} {Influence maximization in complex networks through optimal percolation},\ }\href@noop {} {\bibfield  {journal} {\bibinfo  {journal} {Nature}\ }\textbf {\bibinfo {volume} {524}},\ \bibinfo {pages} {65} (\bibinfo {year} {2015})}\BibitemShut {NoStop}%
\bibitem [{\citenamefont {Morone}\ \emph {et~al.}(2016)\citenamefont {Morone}, \citenamefont {Min}, \citenamefont {Bo}, \citenamefont {Mari},\ and\ \citenamefont {Makse}}]{morone2016collective}%
  \BibitemOpen
  \bibfield  {author} {\bibinfo {author} {\bibfnamefont {F.}~\bibnamefont {Morone}}, \bibinfo {author} {\bibfnamefont {B.}~\bibnamefont {Min}}, \bibinfo {author} {\bibfnamefont {L.}~\bibnamefont {Bo}}, \bibinfo {author} {\bibfnamefont {R.}~\bibnamefont {Mari}},\ and\ \bibinfo {author} {\bibfnamefont {H.~A.}\ \bibnamefont {Makse}},\ }\bibfield  {title} {\bibinfo {title} {Collective influence algorithm to find influencers via optimal percolation in massively large social media},\ }\href@noop {} {\bibfield  {journal} {\bibinfo  {journal} {Scientific reports}\ }\textbf {\bibinfo {volume} {6}},\ \bibinfo {pages} {30062} (\bibinfo {year} {2016})}\BibitemShut {NoStop}%
\bibitem [{\citenamefont {Pei}\ \emph {et~al.}(2018)\citenamefont {Pei}, \citenamefont {Morone},\ and\ \citenamefont {Makse}}]{Pei2018}%
  \BibitemOpen
  \bibfield  {author} {\bibinfo {author} {\bibfnamefont {S.}~\bibnamefont {Pei}}, \bibinfo {author} {\bibfnamefont {F.}~\bibnamefont {Morone}},\ and\ \bibinfo {author} {\bibfnamefont {H.~A.}\ \bibnamefont {Makse}},\ }\bibinfo {title} {Theories for influencer identification in complex networks},\ in\ \href {https://doi.org/10.1007/978-3-319-77332-2_8} {\emph {\bibinfo {booktitle} {Complex Spreading Phenomena in Social Systems: Influence and Contagion in Real-World Social Networks}}},\ \bibinfo {editor} {edited by\ \bibinfo {editor} {\bibfnamefont {S.}~\bibnamefont {Lehmann}}\ and\ \bibinfo {editor} {\bibfnamefont {Y.-Y.}\ \bibnamefont {Ahn}}}\ (\bibinfo  {publisher} {Springer International Publishing},\ \bibinfo {address} {Cham},\ \bibinfo {year} {2018})\ pp.\ \bibinfo {pages} {125--148}\BibitemShut {NoStop}%
\bibitem [{\citenamefont {Moreno}\ \emph {et~al.}(2004)\citenamefont {Moreno}, \citenamefont {Nekovee},\ and\ \citenamefont {Pacheco}}]{moreno2004dynamics}%
  \BibitemOpen
  \bibfield  {author} {\bibinfo {author} {\bibfnamefont {Y.}~\bibnamefont {Moreno}}, \bibinfo {author} {\bibfnamefont {M.}~\bibnamefont {Nekovee}},\ and\ \bibinfo {author} {\bibfnamefont {A.~F.}\ \bibnamefont {Pacheco}},\ }\bibfield  {title} {\bibinfo {title} {Dynamics of rumor spreading in complex networks},\ }\href@noop {} {\bibfield  {journal} {\bibinfo  {journal} {Physical review E}\ }\textbf {\bibinfo {volume} {69}},\ \bibinfo {pages} {066130} (\bibinfo {year} {2004})}\BibitemShut {NoStop}%
\bibitem [{\citenamefont {Zhao}\ \emph {et~al.}(2012)\citenamefont {Zhao}, \citenamefont {Wang}, \citenamefont {Chen}, \citenamefont {Wang}, \citenamefont {Cheng},\ and\ \citenamefont {Cui}}]{zhao2012sihr}%
  \BibitemOpen
  \bibfield  {author} {\bibinfo {author} {\bibfnamefont {L.}~\bibnamefont {Zhao}}, \bibinfo {author} {\bibfnamefont {J.}~\bibnamefont {Wang}}, \bibinfo {author} {\bibfnamefont {Y.}~\bibnamefont {Chen}}, \bibinfo {author} {\bibfnamefont {Q.}~\bibnamefont {Wang}}, \bibinfo {author} {\bibfnamefont {J.}~\bibnamefont {Cheng}},\ and\ \bibinfo {author} {\bibfnamefont {H.}~\bibnamefont {Cui}},\ }\bibfield  {title} {\bibinfo {title} {Sihr rumor spreading model in social networks},\ }\href@noop {} {\bibfield  {journal} {\bibinfo  {journal} {Physica A: Statistical Mechanics and its Applications}\ }\textbf {\bibinfo {volume} {391}},\ \bibinfo {pages} {2444} (\bibinfo {year} {2012})}\BibitemShut {NoStop}%
\bibitem [{\citenamefont {Engsig}\ \emph {et~al.}(2022)\citenamefont {Engsig}, \citenamefont {Tejedor},\ and\ \citenamefont {Moreno}}]{engsig2022}%
  \BibitemOpen
  \bibfield  {author} {\bibinfo {author} {\bibfnamefont {M.}~\bibnamefont {Engsig}}, \bibinfo {author} {\bibfnamefont {A.}~\bibnamefont {Tejedor}},\ and\ \bibinfo {author} {\bibfnamefont {Y.}~\bibnamefont {Moreno}},\ }\bibfield  {title} {\bibinfo {title} {Robustness assessment of complex networks using the idle network},\ }\href {https://doi.org/10.1103/PhysRevResearch.4.L042050} {\bibfield  {journal} {\bibinfo  {journal} {Phys. Rev. Res.}\ }\textbf {\bibinfo {volume} {4}},\ \bibinfo {pages} {L042050} (\bibinfo {year} {2022})}\BibitemShut {NoStop}%
\end{thebibliography}

%

\section*{acknowledgments}
Y.M was partially supported by the Government of Arag\'on, Spain and ``ERDF A way of making Europe'' through grant E36‐23R (FENOL), and by Ministerio de Ciencia e Innovaci\'on, Agencia Espa\~nola de Investigaci\'on (MCIN/AEI/10.13039/501100011033) Grant No. PID2020‐115800GB‐I00. A.T. thanks the Spanish Ministry of Universities and the European Union Next Generation EU/PRTR for their support through the Maria Zambrano program. A.T. and E.F-G. were partially
supported by NSF (EAR1811909) and the United Kingdom Research \& Innovation Living Deltas Hub NES008926.  E.F.-G. also acknowledges partial support by NASA (Grants 80NSSC22K0597 and 80NSSC23K1304) and NSF (Grant IIS2324008). The funders had no role in the study design, data collection, analysis, the decision to publish, or the preparation of the manuscript.

\section*{Contributions}
M.E and A.T developed the methods. M.E. implemented the code and conducted the experiments. M.E and A.T analyzed the data. M.E. and A.T. wrote the initial manuscript. All authors discussed the results, revised the manuscript, and approved the final version of the manuscript.

\section*{Competing interests}
The authors declare no competing interests.

\end{document}


\preprint{}

\title{Supplementary Material: \\DomiRank Centrality: revealing structural fragility of \\ complex networks via local dominance}

\author{Marcus Engsig}
 \email{marcus.w.engsig@gmail.com}

\affiliation{%
Directed Energy Research Centre, Technology Innovation Institute, Abu Dhabi, United Arab Emirates.
}%

\author{Alejandro Tejedor}%
 \email{alej.tejedor@gmail.com}
\affiliation{%
 Institute for Biocomputation and Physics of Complex Systems (BIFI), Universidad de Zaragoza, 50018 Zaragoza, Spain
}%
\affiliation{
 Department of Theoretical Physics, University of Zaragoza, Zaragoza 50009, Spain
}%

\affiliation{%
 Department of Civil and Environmental Engineering, University of California Irvine, Irvine, CA 92697, USA.
}%

\author{Yamir Moreno}
\email{yamir.moreno@gmail.com}
\affiliation{%
 Institute for Biocomputation and Physics of Complex Systems (BIFI), University of Zaragoza, 50018 Zaragoza, Spain
}%
\affiliation{
 Department of Theoretical Physics, University of Zaragoza, Zaragoza 50009, Spain
}%
\affiliation{CENTAI Institute, Turin 10138, Italy.
}%

\author{Efi Foufoula-Georgiou}
\email{efi@uci.edu}
\affiliation{%
 Department of Civil and Environmental Engineering, University of California Irvine, Irvine, CA 92697, USA.
}%
\affiliation{
 Department of Earth System  Science, University of California Irvine, Irvine, CA 92697, USA.
}%

\author{Chaouki Kasmi}
\email{chaouki.kasmi@tii.ae}
\affiliation{%
Directed Energy Research Centre, Technology Innovation Institute, Abu Dhabi, United Arab Emirates
}%

\maketitle

\tableofcontents
\newpage


\section{Understanding power shifts in Rich-Clubs using DomiRank}
 In this section, we present a first exploration of Rich-Club networks via DomiRank to provide insight into (i) the joint dominance (collusion) mechanism that emerges from competition dynamics and (ii) how changing the connectivity of the Rich Club nodes and/or their peripheries alters in an interpretable manner the relative dominance of Rich-Club's individuals and therefore, the overall power dynamics of Rich-Club networks.

Rich-Club networks are important for understanding various types of real networks - i.e., political networks, macroscale brain networks, and many more \cite{colizza2006detecting, mcauley2007rich, van2011rich, Alstott2014}. For these networks, commonly used centralities have difficulty distinguishing the importance between nodes within a Rich Club, as their nodes establish a closely related `hierarchy' of hubs. Here, DomiRank offers a  novel insight to reveal key properties of the structure and dynamics of Rich-Club networks. For this section, we explore a simple version of a Rich-Club network, which consists of an interconnected Rich Club of hub nodes. Each hub node in the Rich Club is a star network with a varying number or nodes attached to them (periphery). Note that we keep the total size of the network sufficiently small to enhance interpretability and visually guide the understanding of the underlying dynamics in terms of DomiRank. \\

\begin{figure*}
    \centering
    \includegraphics[width = \linewidth]{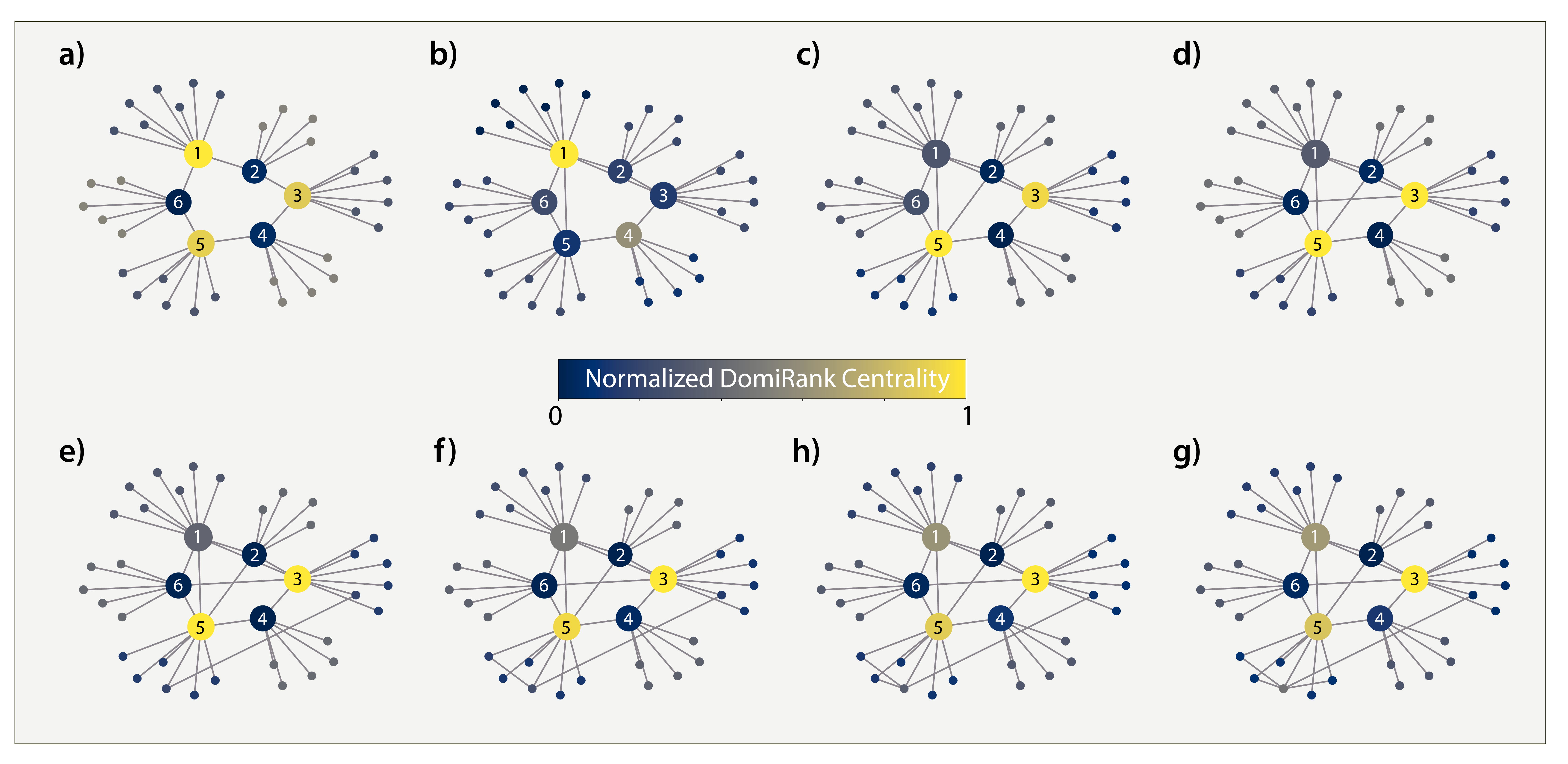}
    \caption{\textbf{Steering dominance in a Rich-Club network by altering connectivity}. The different panels show changes in dominance (color-encoded) starting from a (a) simple Rich-Club network when new competition (links) is established within the Rich-Club and/or in its periphery (note that some nodes are arbitrarily labeled for reference). In particular, the following new links were introduced in each of the panels: (b) link (1,3) and link (1,5); (c) link (2,5); (d) link (3,6); (e) link between the periphery of node 3 and the periphery of node 5; and (f,g,h) internal competition (addition of links) in the periphery of node 5. In all the panels, the DomiRank centralities are computed for $\sigma = \frac{0.99}{-\lambda_N}$ in order to simulate a highly competitive environment where changes in power dynamics are more likely to occur.
    }
    \label{fig:SM_RC_1}
\end{figure*}

Fig. \ref{fig:SM_RC_1} illustrates how DomiRank centrality can be instrumental in providing fundamental understanding of the power dynamics on Rich-Club networks characterized by highly competitive relations. Fig. \ref{fig:SM_RC_1}a displays the chosen initial configuration of the Rich-Club, where node coloring represents nodal dominance given by DomiRank centrality  ($\sigma = \frac{0.99}{-\lambda_N}$), and nodes are arbitrarily labeled with numbers to ease the discussion that follows. Here, node $1$ is dominating (highest DomiRank score) as it has the largest periphery, and additionally, it is jointly dominating nodes $2$ and $6$, by colluding with nodes $3$ and $5$, respectively. Thus, nodes $3$ and $5$ exhibit dominant behavior, although below the levels obtained by node $1$.  Actually, node $1$ can reinforce its dominant role by establishing direct competition (links) with nodes $3$ and $5$ (Fig. \ref{fig:SM_RC_1}b). In such a scenario, node $1$ is able to subdue nodes $3$ and $5$ with the help of node $4$, which contributes via joint dominance.  A more drastic power shift such that nodes $3$ and $5$ jointly dominate the rich club can also be induced by establishing competition between nodes $5$ and $2$, and $3$ and $6$ (Fig. \ref{fig:SM_RC_1}c,d respectively).

Furthermore, changes in the relative dominance of the nodes in the rich club can be promoted by modifying the connectivity of their peripheries.  Thus, the dominance of node $3$ over node $5$ can be induced by creating competition in the periphery of node $5$, which would internally challenge the dominance of node $5$ (Fig. \ref{fig:SM_RC_1}e-g). In other words, an increase in the connectivity (competition) of the periphery reduces the relative dominance of the center node connected to that periphery (node 5), making its periphery less dominated and more `fit'. The power shift observed from introducing competition in the periphery of a hub is not as drastic as that of creating direct competition for members within the Rich Club; however, it is sufficient to steer the power balance within the Rich Club.

These simple examples show how DomiRank is able to naturally describe collusion scenarios where nodes can dominate neighborhoods and gain power by establishing common enemies. Therefore, we show that we can steer power dynamics by artificially forcing competition and joint dominance between Rich-Club nodes or by encouraging competition within a hub periphery.

\section{DomiRank and strict symmetry constraints: even Lattice}
As demonstrated by the results shown in the main text, DomiRank is able to integrate, for large values of $\sigma$, global (meso- to large-scale) network properties. This is particularly apparent in the case of a square lattice network with an odd number of nodes (see Fig. 2 in the main text), where the entire network properties partially influence the value of each node's DomiRank via the competition mechanism. At the limit of high $\sigma$  an alternating spatial pattern emerges which reflects two global network properties: the finite boundary and global symmetries.

\begin{figure*}
    \centering
    \includegraphics[width = \linewidth]{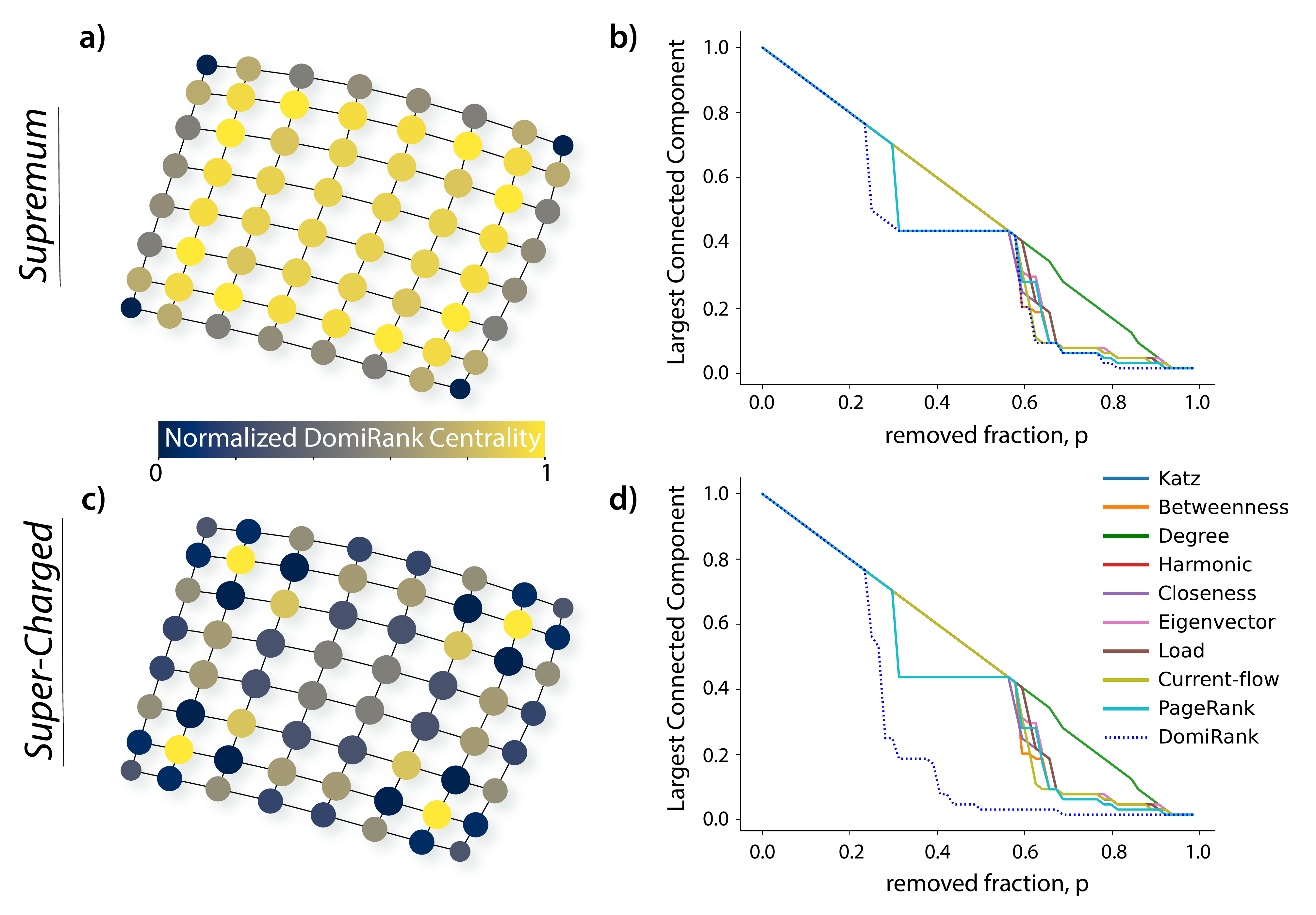}
    \caption{\textbf{Supercharging DomiRank competition on an even lattice}. DomiRank centrality distributions  in an even lattice of size $N=8\times 8$ for competition levels corresponding to (a) $\sigma = \frac{0.99}{-\lambda_N}$ and (c) $\sigma = \frac{1.25}{-\lambda_N}$. Panels (b,d) show the deterioration in the relative size of the largest connected component according to the attack strategies based on the centrality values shown in panels (a,c), respectively, along with nine other centralitiy-based attacks.}
    \label{fig:SM_even_lattice}
\end{figure*}

However, in rare cases, strict symmetries can exert constraints that hinder the emergence of a fully developed dominance pattern. A prime example of such a scenario corresponds to a square lattice with an even number of nodes (see Fig. \ref{fig:SM_even_lattice}a). We show that in such a network, due to the degree homogeneity and strict symmetry, DomiRank cannot establish an alternating pattern of dominating-dominated nodes similar to the one exhibited by the square lattice with an odd number of nodes, as it would break the system's symmetries. Thus, for instance, the spatial pattern of DomiRank obtained for a square lattice of 64 nodes is shown in Fig. \ref{fig:SM_even_lattice}a at the limit $\sigma \to \frac{1}{-\lambda_N}$. We note that for the DomiRank distribution based on this value of $\sigma$, attacks are still more efficient than those based on different centralities (see Fig. \ref{fig:SM_even_lattice}b). However, the overall efficiency is significantly lower than for the DomiRank-based attack for a square lattice with an odd number of nodes, as its superior efficiency relies on the alternating nature of the spatial distribution.  Interestingly, we have found numerically that in the case of the lattice with an even number of nodes, the optimal value of the parameter corresponds to $\sigma_{opt} = \frac{1.25}{-\lambda_N}$, i.e.,  a case of supercharged competition. Note that for this network, despite $\sigma$ being outside of the interval  $\left(0, \frac{1}{-\lambda_N}\right)$, the recursive formula still yields numerical convergence (to tolerance levels of at least  $10^{-6}$).  For this state of supercharged competition, an alternating spatial pattern of DomiRank is obtained (Fig. \ref{fig:SM_even_lattice}c), which, although different from that displayed for the lattice with an odd number of nodes, is able to produce comparably efficient attacks with the odd case (Fig. \ref{fig:SM_even_lattice}d). When a similar supercharged competition was explored in other topologies, the analytical DomiRank calculation led to suboptimal attack strategies, and numerical divergence via the recurrence equation was obtained.

\section{Finding Optimal $\sigma$}
DomiRank has a tuneable parameter $\sigma$, that allows modulation of the level of competition introduced in the dynamical equation. In order to find the optimal $\sigma$, we first define our loss function - the area under the largest-connected-component curve generated from a network undergoing sequential node removal - and thereafter numerically explore the loss function for the full range of $\sigma$ values. Choosing the $\sigma$ that corresponds to the smallest loss, yields the optimal $\sigma^*$ in terms of efficiently dismantling the network.

In this section, we show the loss curves for varying $\sigma$ for various topologies - both real and synthetic. The loss curves are displayed for two main reasons: (i) to show the typical behavior that they exhibit for different topologies, and (ii) to demonstrate that this methodology is applicable to massive networks, as the space can be searched for a network size of nearly 24 million nodes.

\begin{figure*}[t]
    \centering
    \includegraphics[width = \linewidth]{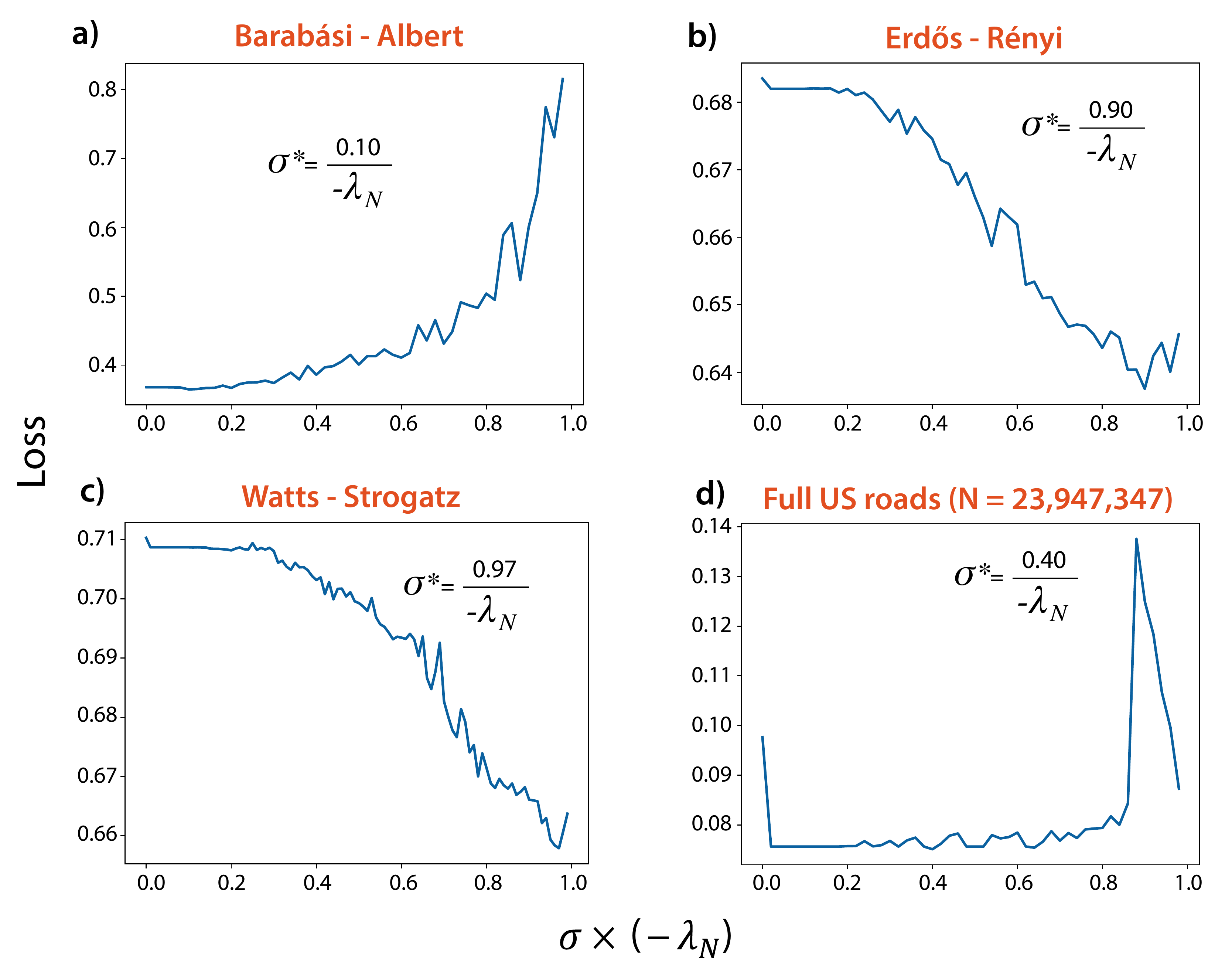}
    \caption{\textbf{DomiRank's loss functions for different topologies.} The DomiRank's loss functions computed for the complete range of the free parameter ($\sigma$) are shown for three different synthetic networks of size $N=1000$: (a) Bar{\'a}basi-Albert, (b) Watts-Strogatz, and (c) Erd\H{o}s-R\'enyi, and for (d) a massive real network, the Full US Road network, consisting of nearly 24 million nodes.  The optimal values of $\sigma^*$ corresponding to the locations of the loss function's global minima are displayed for each panel. Note that all the curves shown were computed using $100$ samples linearly distributed in the interval $ (0, \frac{1}{-\lambda_N})$. 
    }
    \label{fig:SM_loss}
\end{figure*}

In Fig. \ref{fig:SM_loss}, we show how some demonstrative loss curves look for different synthetic topologies and for a real-world network (the full US roads network) consisting of $N = 23,947,347$ nodes. The optimal level of competition ($\sigma^*$) is displayed in each panel. We want to highlight a few important observations from Fig. \ref{fig:SM_loss}: (i) Loss curves exhibit clear decreasing and/or increasing trends towards the area of optimal $\sigma$ (with some variability);  (ii) For many topologies, there is a range of values of $\sigma$ for which the attack is near optimal; And (iii) different topologies require a different trade-off of local vs. global information ($\sigma$) in order to dismantle the network most efficiently. Finally, Fig. \ref{fig:SM_loss}d shows that the optimal parameter can also be found for massive networks. 

Regarding the computational cost of exploring the optimal value of $\sigma$, we first need to consider that the cost of computing the connected components of a graph scales with $\mathcal{O}(N + m)$ \cite{pearce2005improved}. Secondly, we can recompute the largest-connected-component size not after every node removal but sample the curve at a given frequency - e.g. set it to be recomputed every $1\%$ of nodes removed. Therefore, if we assume that links are removed linearly (which is near worst case - see Fig. \ref{fig:SMLinks}), then the computational cost for dismantling the entire network and recomputing the largest connected component every $1\%$ ($100$ times) scales with $\mathcal{O}((\frac{100+1}{2}(N+m))$. Finally, computing the area under the largest connected component for different $\sigma$ are independent calculations, and thus, this loss function as a function of $\sigma$ can be parallelized, and sped up almost linearly with the number of threads.

\begin{figure*}[t]
    \centering
    \includegraphics[width = \linewidth]{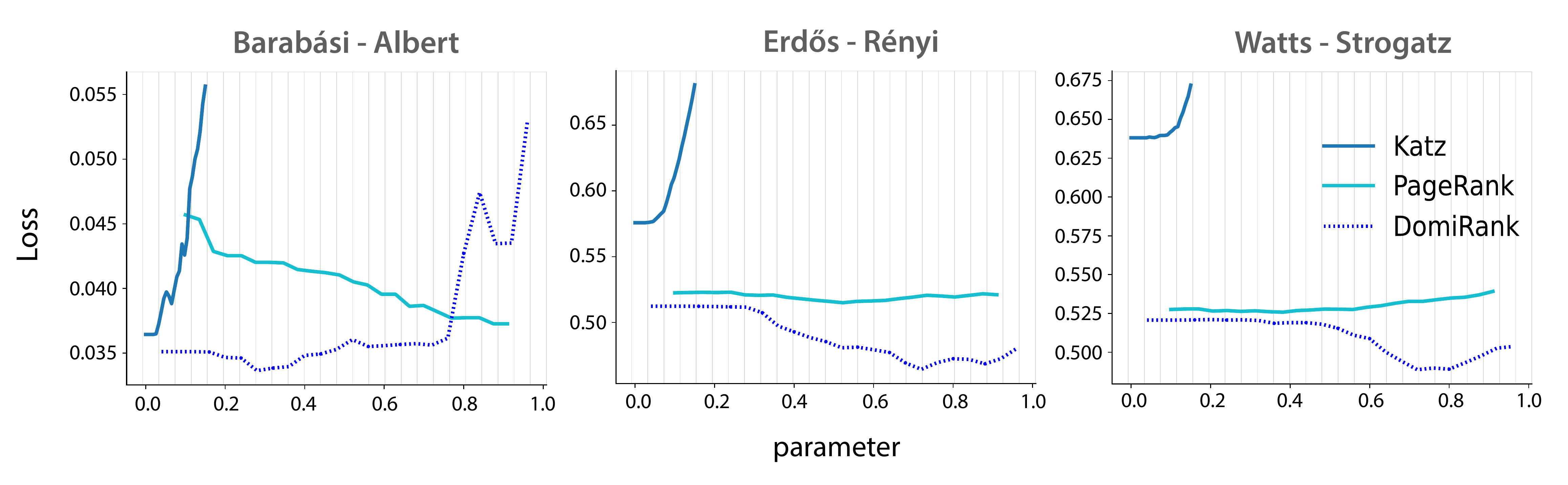}
    \caption{\textbf{Comparing the loss functions of PageRank, Katz, and DomiRank within their parameter space.} The loss functions corresponding to PageRank, Katz, and DomiRank computed for the complete range of their free parameters are shown for three different synthetic networks of size $N=300$, namely, (a) Bar{\'a}basi-Albert, (b) Erd\H{o}s-R\'enyi, and (c) Watts-Strogatz. Note that all the curves shown were computed using $25$ samples linearly distributed in their respective parameter domains.}
    \label{fig:SM_katzPageRank}
\end{figure*}

Note that other centrality metrics also have a parameter in their definition. In those cases, we have used their standard values for our analysis. Thus, Katz's and PageRank's parameters were chosen to be $0.01$ and $0.85$, respectively. The Katz's parameter was reduced to $0.001$ occasionally, when convergence was not found within $1000$ iterations. We have further tested that this election is not detrimental to their performance in comparison to DomiRank and other metrics. Fig. \ref{fig:SM_katzPageRank} shows that the gain from choosing the optimal value for the Katz and PageRank parameters is marginal when compared to their default values. Finally, we also show that DomiRank tends to outperform the other metrics, regardless of the $\sigma$ parameter.

\section{Correlations of DomiRank with other centralities}
Here, we look at various pairwise Spearman correlations between the centralities tested on four synthetic topologies, namely, (i) Bar{\'a}basi-Albert, (ii) Erd\H{o}s-R\'enyi, (iii) Random Geometric Graph, (iv) and a 2D lattice, in order to gain insights into how DomiRank dismantles a network.

The DomiRank centrality is derived as the steady-state solution of a competition-based dynamical equation, yielding nodal scores inherently different from previous methods. The Spearman correlation provides a good tool to understand how similar DomiRank is to other previous centrality methods. Conceptually, we know from the dynamical equation that as $\sigma \to 0$, DomiRank, by definition converges to the degree distribution, and as $\sigma \to \frac{1}{-\lambda_N}$ an increasing amount of competition (mesoscale and global information) is considered in the dynamics and that adjacent nodes tend to have disparate scores, which, is a quite distinctive signature of DomiRank with regards to other centrality measures. \\

\begin{figure*}
    \centering
    \includegraphics[width = \linewidth]{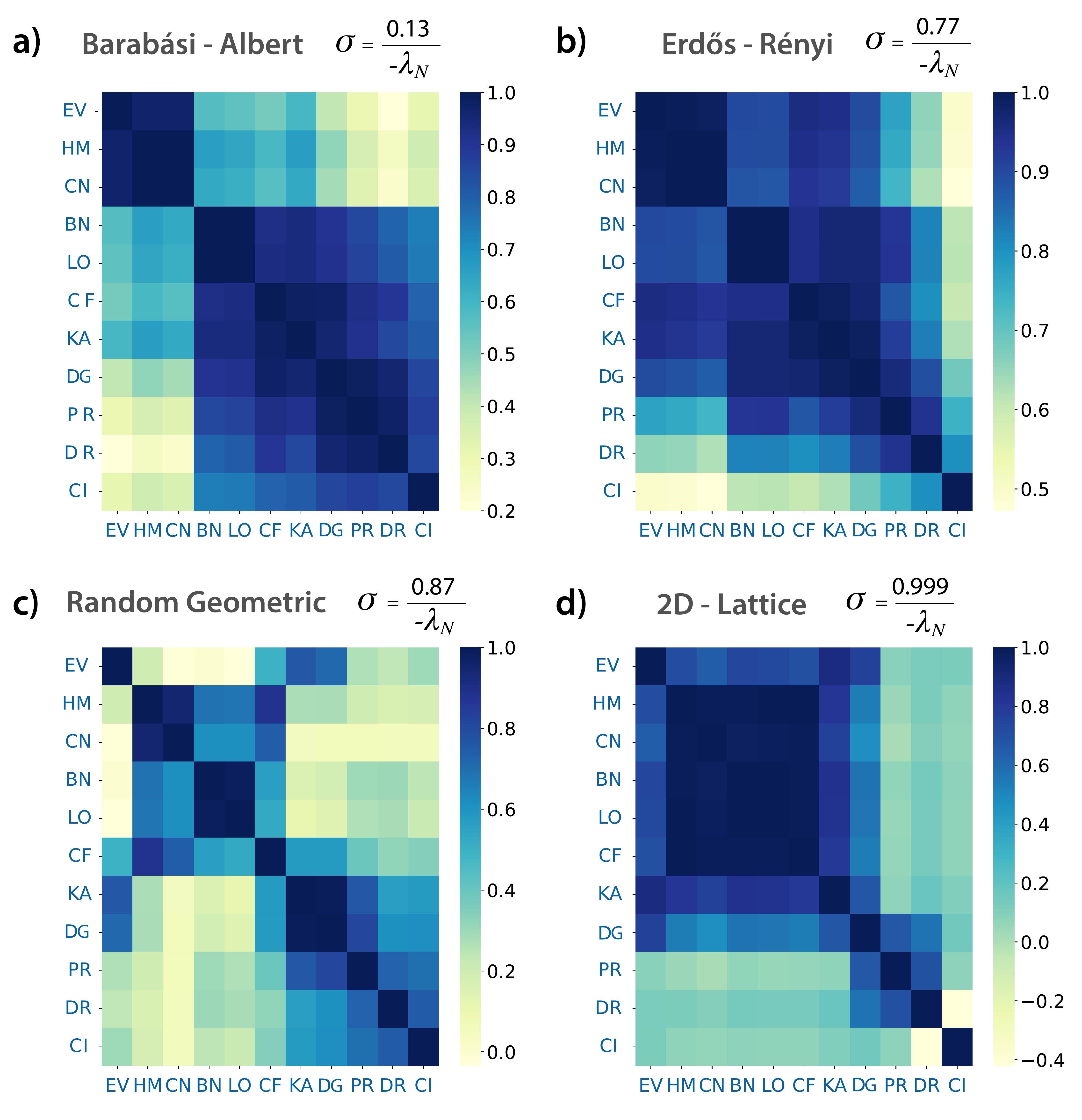}
    \caption{\textbf{Pairwise Spearman correlation between different centralities.} The pairwise Spearman correlation matrices for the different centralities used in our study are displayed as heatmaps for four different topologies, namely, (a) Bar{\'a}basi Albert ($N=1000$), (b) Erd\H{o}s-R\'enyi ($N=1000$), (c) Random Geometric Graph ($N=1000$), and a regular 2D-Lattice ($N=33\times 33$). Particularly, the centralities compared are Eigenvector (EV), Harmonic (HM), Closeness (CN), Betweenness (BN), Load (LO), Current-Flow (CF), Katz (KA), Degree (DG), PageRank (PR), DomiRank (DR), and Collective Influence (CI). 
    }
    \label{fig:SM_spearman}
\end{figure*}

Fig. \ref{fig:SM_spearman}(a-d) presents the pairwise Spearman correlation for several centralities included in our analysis. As noted before, the optimal value of $\sigma$ depends on the topology considered. The topologies analyzed allow us to explore a range of $\sigma$ increasing progressively from Bar{\'a}basi-Albert to the 2D-lattice  (Fig. \ref{fig:SM_spearman}(a-d)). Within the gradient of $\sigma$, we observe the expected progressive divergence of DomiRank from degree centrality. We also observe that DomiRank displays different degrees of correlation to different centralities depending on the topology examined, providing further evidence of the adaptability of DomiRank by varying the value of $\sigma$ to mine the most relevant structural features of each topology. 

For its significance, we have also included a comparison of DomiRank with Collective Influence (CI) \cite{morone2015influence}, despite the latter being the only centrality in this analysis relying on an iterative computation (more details see section S-VI). Interestingly, DomiRank offers high values of Spearman correlation with CI, particularly as $\sigma$ increases (see Fig. \ref{fig:SM_spearman}a-c). This observation is compatible with the fact that both metrics can extract global information to inform the individual nodal centrality and avoid artificially high values for neighboring nodes (although through very different mechanisms  - see section VI for more details).  It might seem striking at first glance that for the case of the 2D-lattice, the Spearman correlation between CI and DomiRank is the lowest among metrics despite the large value of the $\sigma$ parameter in that case. However, this example should not be taken as evidence of the divergence of the two metrics for very large values of $\sigma$; actually, a closer examination of those two metrics on the  2D-Lattice (see Section S-VI) reveals that they differ in the spatial pattern due to the strict symmetry of the network, but they still exhibit similar strategies creating alternating patterns of high-low values.

\section{Heterogeneous networks}
Figure 5 in the main text shows the evolution of the largest connected component for different synthetic and real-world networks as they are attacked based on various centrality metrics. DomiRank-based attacks outperform (with different degrees of significance) all other attacks for all the networks analyzed but in one case. That case corresponds to a massive social network (LiveJournal users and their connections - see Figure 4k in the main text), where the DomiRank-based attack, although very competitive, does not perform better than the PageRank-based attack. 

In this section, we pose the hypothesis that the presence of heterogeneity (different structural properties) in different subgraphs of the network could lead to underperformance for DomiRank-based attacks.  The rationale behind this statement is straightforward in the light of previous results. As shown in the analysis of synthetic networks in Fig. 3 and Fig. 4 of the main text, DomiRank excels in highlighting the important nodes with different values of $\sigma$ depending on the network structure. Thus, for hub-dominated networks, a relatively low value of $\sigma$ provides the most effective ordering of nodes for designing attack strategies. On the other hand, more regular networks such as lattices or random graphs require large values of $\sigma$, which allow for a larger integration of the information in the network structure for assessing the relative importance of each node. Consequently, if a network consists of different subgraphs (e.g., communities) with different topological properties might require different $\sigma$ for each subgraph since an {\it average} global value of $\sigma$ would lead to suboptimal results.

\begin{figure*}[!h]
    \centering
    \includegraphics[width = \linewidth]{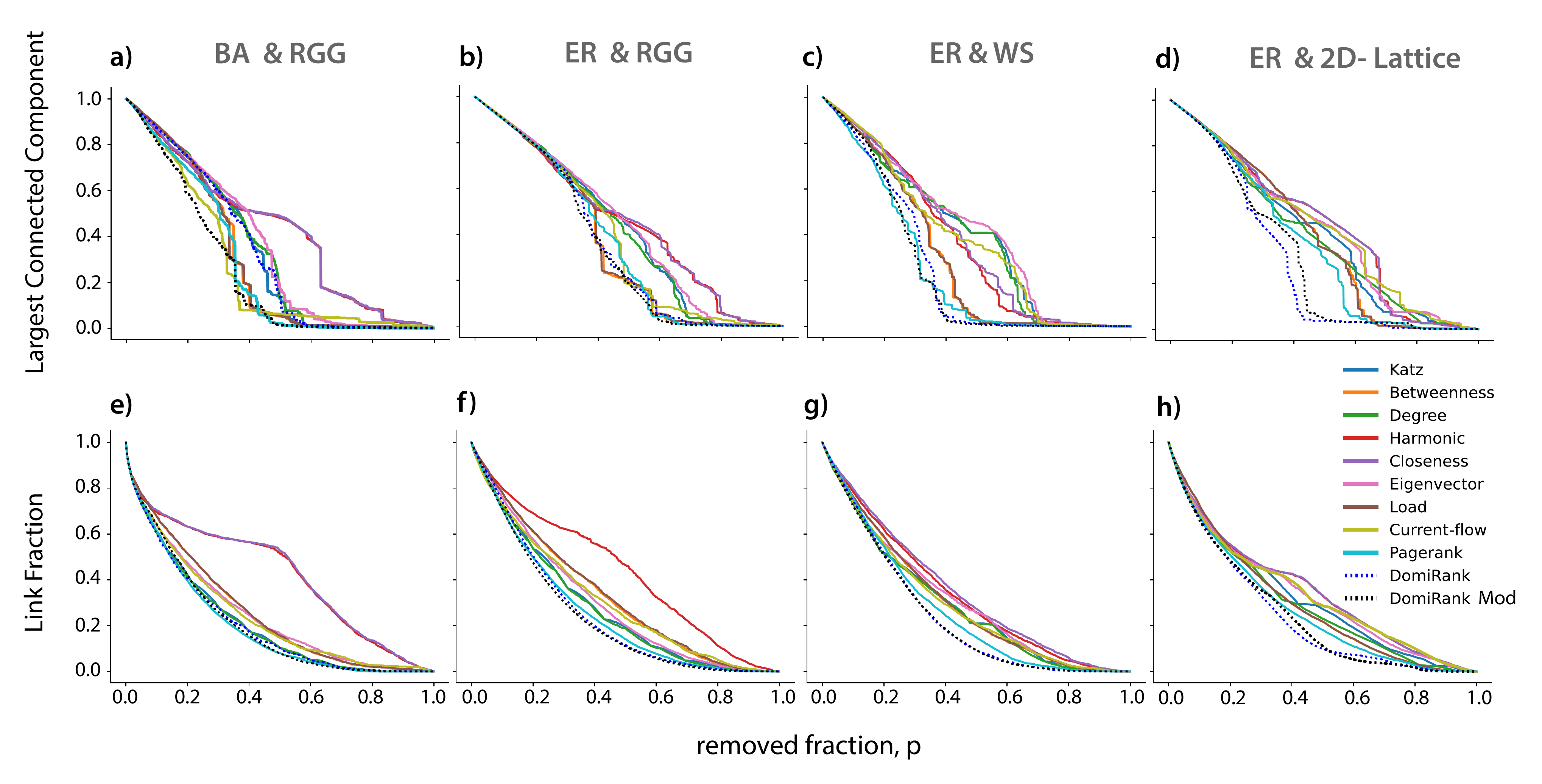}
    \caption{\textbf{The effect of heterogeneity on the performance of centrality-based attacks on synthetic networks.} Panels a-d show the evolution of the relative size of the largest connected component and panels e-h show the evolution of the remaining link fraction whilst undergoing sequential node removal according to descending scores of various centrality measures for different coupled synthetic networks (heterogeneous) of size $N=1000$: (a,e) Bar{\'a}basi-Albert (BA) and random geometric graph (RGG), (b,f) Erd\H{o}s-R\'enyi (ER) and random geometric graph (RGG), (c,g) Erd\H{o}s-R\'enyi (ER)  and Watts-Strogatz (WS), and (d,h) Erd\H{o}s-R\'enyi (ER)  and a 2D-lattice. Here we have two different DomiRank-based-attacks corresponding to two different $\sigma$ (i) the optimal $\sigma$ for the entire network (denoted in the legend as DomiRank), and (ii) the capped optimal diagonal matrix $\sigma$ as per eq. \ref{eq:diagonalSigmaClipped} (denoted in the legend as DomiRank Mod).
    }
    \label{fig:SMHeterogeneity}
\end{figure*}

In order to address this issue, we could substitute the $\sigma$ parameter in DomiRank definition by a diagonal matrix (without additional computational cost), where the entries of $\sigma_{i,i}$ are corresponding to the optimal $\sigma$ for the community that node $i$ belongs to. We can mathematically describe this new diagonal matrix $\sigma$ as follows, given $T$ communities $C_j, j \in [1,T]$;
\begin{equation}
    (\sigma)_{i,i} = \sum_{j=1}^T \sigma_j \mathds{1}_{ i\in C_j}.
    \label{eq:diagonalSigma}
\end{equation}
Moreover, in order to guarantee the convergence of DomiRank, Eq. \ref{eq:diagonalSigma} takes the final form:

\begin{equation}
(\sigma)_{i,i} = \min \left[ \sum_{j=1}^T \sigma_j\mathds{1}_{ i\in C_j}, \frac{-1}{\lambda_N} \right]
   \label{eq:diagonalSigmaClipped}
\end{equation}

\noindent where $\lambda_N$ is the minimum (largest negative) eigenvalue of the whole network.

To test this hypothesis, we generate synthetic networks consisting of two subgraphs, each of them generated with a different model (e.g., Bar{\'a}basi-Albert and random geometric graph), establishing links between both subgraphs ($10\%$ of the nodes establish a connection with a node in the other subgraph). Fig. \ref{fig:SMHeterogeneity} displays the results for the attacks based on the previous centralities (including DomiRank), as well as the results obtained from a DomiRank where nodes in different subgraphs are evaluated with different values of $\sigma$  to account for heterogeneity in networks. As expected, in the cases where the two merged networks are characterized by disparate values of $\sigma$ for optimal attack strategies (i.e., large heterogeneity),  we obtain a larger gain by considering the diagonal-matrix-based $\sigma$ in the definition of DomiRank. Thus, for instance, Fig. \ref{fig:SMHeterogeneity}a shows a significant improvement in the performance of the attack strategy when heterogeneity is considered in the definition of DomiRank for a network consisting in the combination of a subgraph generated by a Bar{\'a}basi-Albert model with relatively low degree ($\bar{k} = 4 $) and random-geometric-graph ($\bar{k} = 5$). Additionally, from Fig. \ref{fig:SMHeterogeneity}c we see that by combining two subgraphs characterized by comparable optimal $\sigma$ values like the Erd\H{o}s-R\'enyi  and Watts Strogatz networks, the gain is just incremental. Note that when the difference in the optimal value of $\sigma$ does not significantly differ between the two subgraphs (e.g., Erd\H{o}s-R\'enyi and 2D-Lattice - See fig. \ref{fig:SMHeterogeneity} d), the traditional DomiRank computed in the whole network might offer better performance than the community-based version, as it accounts for the links connecting the two sub-graphs. 

Consequently, massive networks consisting of multiple communities with different properties might require the adoption of the definition of DomiRank that accounts for that heterogeneity (i.e., using the diagonal-matrix-based $\sigma$) to design more efficient attack strategies for dismantling the networks, as the traditional definition of DomiRank could underperform. 

Thus, by combining the various algorithms to detect communities in massive networks, and this newly defined $\sigma$ could potentially lead to further gains in designing strategies to dismantle networks' structure and functionality without incurring unaffordable computational costs.

\section{Collective Influence (CI) and DomiRank} 
In this section, Collective Influence (CI) and DomiRank are compared side-by-side, to understand their differences in terms of (i) underlying principles and (ii) ability to dismantle toy networks. 

CI, a relatively new centrality metric introduced by Morone and Makse \cite{morone2015influence}, is defined for a node  $i$ as follows,
\begin{equation}
    CI_\ell(i) = (k_i -1 ) \sum_{j \in \partial B(i,\ell)} (k_j - 1),
    \label{eq:SM_CI}
\end{equation}
where, $\ell$ is a parameter, $\partial B(i, \ell)$ is the frontier of the ball containing the nodes around node $i$ at a distance $\ell$ (without back-propagation), and $k_{i}, k_j$ are the degrees of node $i,j$ respectively. CI has a low computational cost, but it is an iterative centrality: once the CI is computed for all the nodes in the network, the node with the highest CI value is formally assigned this value and removed from the network. Then, the CI is recomputed for all the remaining nodes, repeating the process of assignation and removal until all nodes are removed.

CI measures the multiplicative cascade of paths that each node could develop at various scales through a tuneable parameter. Morone and Makse \cite{morone2015influence} showed that collective influence found novel influencers in networks that had remained unseen by previous centrality measures \cite{morone2016collective}. These novel influencers were identified to be nodes surrounded by hierarchical coronas of hubs, making them potential epicenters for spreading processes. In our study, we follow the methodology of Pei et al. \cite{Pei2018}, and set $\ell = 2$. 

On the other hand, DomiRank identifies the relative dominance of the different nodes in their respective neighborhoods, capitalizing on a dynamical equation that incorporates a competition term and a relaxation term.  For ease of comparison, DomiRank's underlying dynamical equation, described in detail in the main text, is displayed here:
\begin{equation}
    \frac{d\Gamma_i(t)}{dt} = \sigma \sum_{j=1}^N [A_{ij}(1 - \Gamma_j(t))] - \Gamma_i(t),
    \label{eq:SM_DomiRank}
\end{equation}
\noindent where $\sigma$ (tuneable parameter) modulates the relative extent of competition in the network and, thus, the amount of global information considered by DomiRank. Note that in this version of the equation, other parameters ($\beta$ and $\theta$) of the more general dynamical equation (see Eq. 1 in the main text) that do not impact the final relative value of the DomiRank distribution are considered equal to one for simplicity. 

Both CI and DomiRank are able to provide global information that is not redundant in neighboring nodes. However, the ways they extract this global information and avoid artificial local correlations are entirely different. Thus, global information is integrated by DomiRank via a competition dynamic, while collective influence accounts for the multiplicative cascade of paths that each node could develop at various scales. In terms of avoiding spurious correlations among neighboring nodes, CI requires an iterative recomputation wherein the most influential node is iteratively removed between each computation of CI, whereas DomiRank is computed only once for the entire network, and it is its underlying dynamic process that is able to inherently provide centrality scores that are not locally redundant. In order to further understand the differences and similarities between DomiRank and CI, we compare the two centralities for small toy networks.

\subsection{Toy Networks}
\begin{figure*}[!h]
    \centering
    \includegraphics[width = \linewidth]{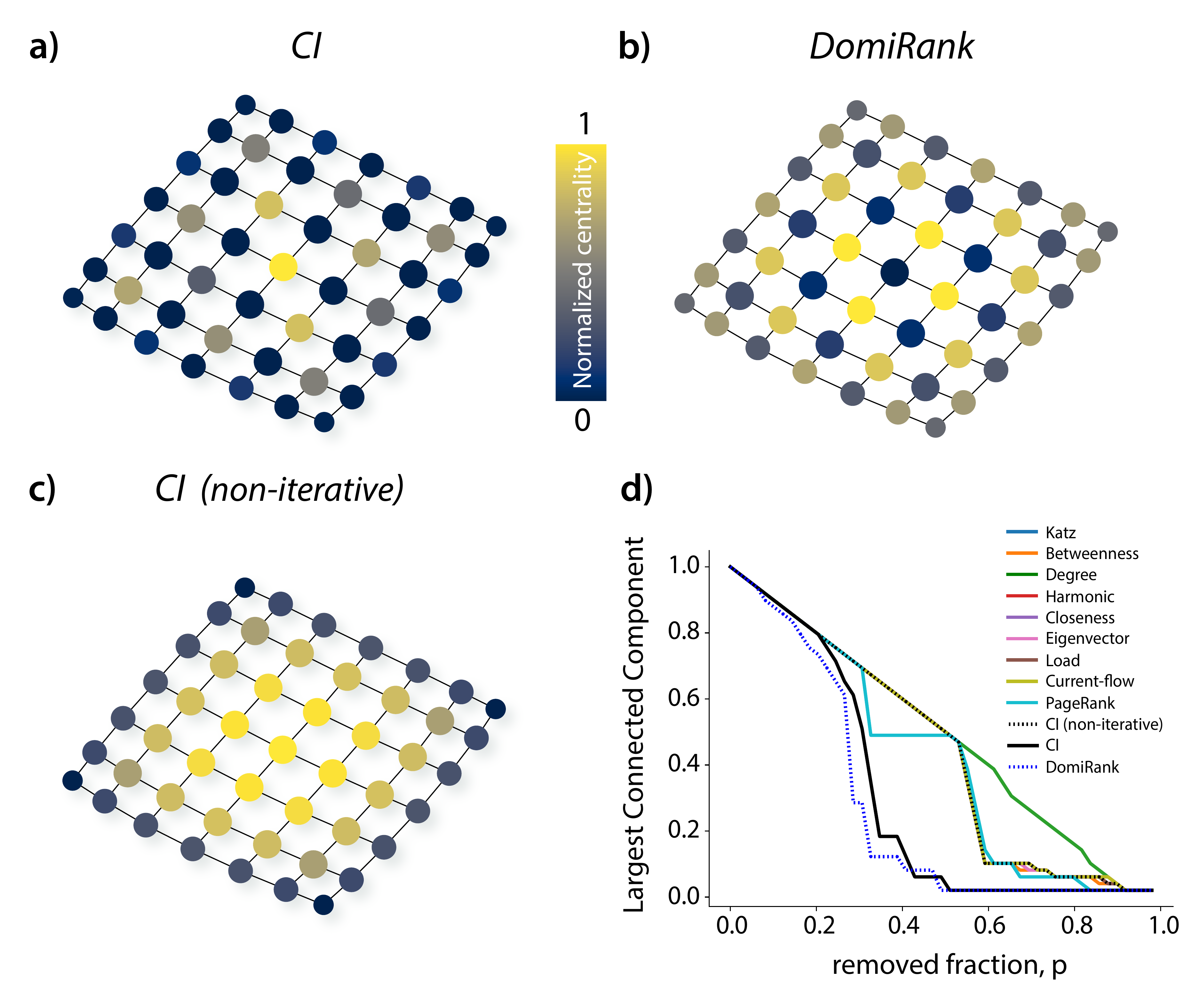}
    \caption{\textbf{Collective Influence and DomiRank on a regular 2D lattice.} Comparison between three centralities on a 2D regular lattice $(N=49)$, namely, (a) Collective Influence, (b) DomiRank, and (c) non-iterative Collective Influence, along with (d) the evolution of the relative size of the largest connected component whilst undergoing sequential node removals according to the displayed centralities.}
    \label{fig:SM_DRCI_ToyGRID}
\end{figure*}

A particularly insightful example for highlighting the differences between DomiRank and CI is the 2D lattice network, which has two key properties - (i) degree-homogeneity apart from the edge of the lattice, and (ii) ordered structure with high symmetry. Fig. \ref{fig:SM_DRCI_ToyGRID}a,b reveals that both centrality distributions exhibit alternating spatial patterns in the network domain. We want to highlight once more here that those alternating patterns emerging for both metrics respond to completely different mechanisms - while it is a direct consequence of the dynamics of DomiRank for highly competitive environments, it is not intrinsic to the definition of CI (see \ref{fig:SM_DRCI_ToyGRID}c for the initial evaluation of CI for every node) but it is the effect of the iterative recomputation of CI after each node removal.

A closer look at Fig. \ref{fig:SM_DRCI_ToyGRID}a,b shows that the two spatial patterns are inverted with respect to each other (see also Fig. \ref{fig:SM_spearman}d for correlations). These antagonistic patterns could be distilled down to the identification of the most important node in the network (as this determines the alternating spatial pattern given the network symmetry). CI finds that the central node is the most important node for a multiplicative cascade of paths, and DomiRank finds that the central node is the most dominated node through the collusion of its neighbors. DomiRank, rather than removing the central node, isolates it, impeding as well its potential in a spreading process whilst continuously damaging the network. Note that both strategies yield very efficient attack strategies, but the highlighted differences between DomiRank and CI lead to slightly better performance for DomiRank-based attacks, as shown by Fig. \ref{fig:SM_DRCI_ToyGRID}d. 

\begin{figure*}[!h]
    \centering
    \includegraphics[width = \linewidth]{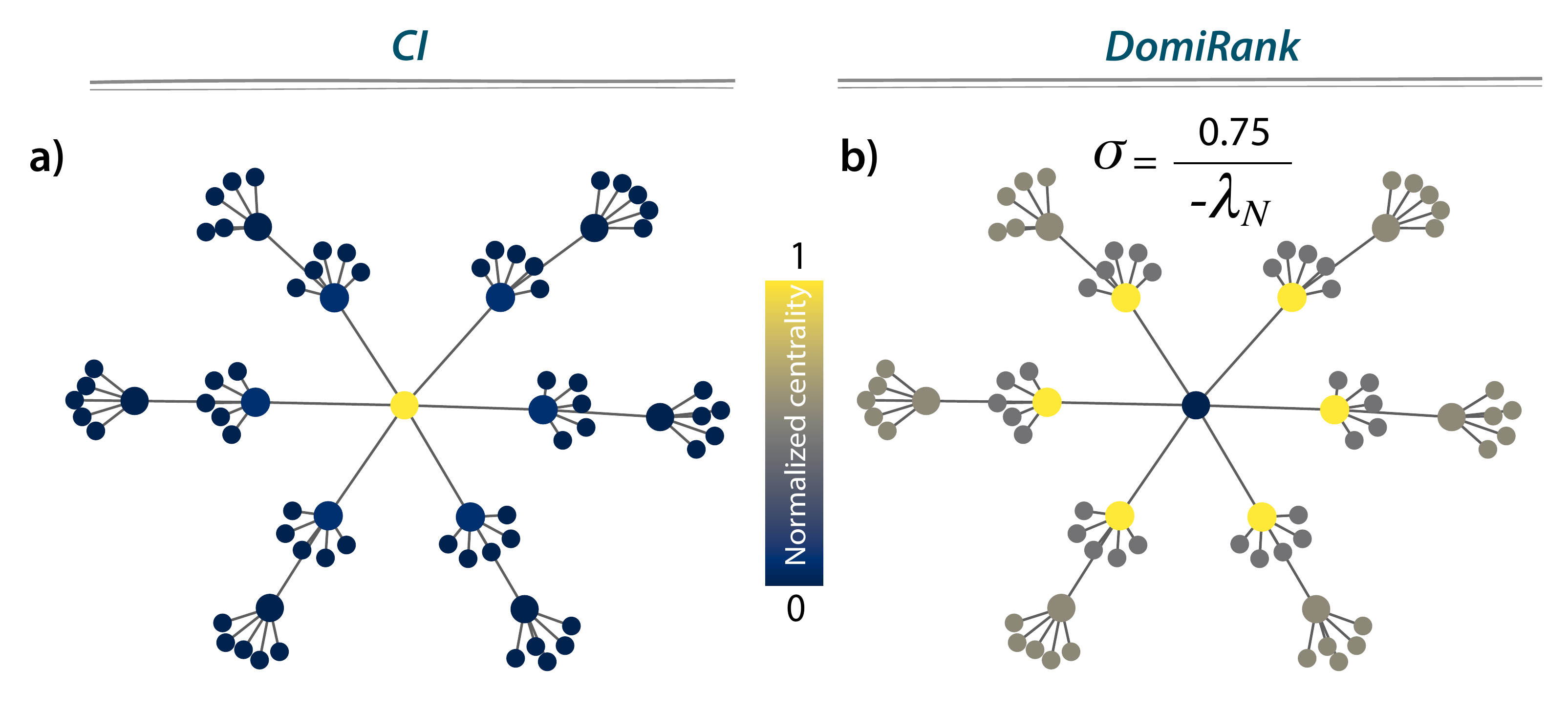}
    \caption{\textbf{Comparison of Collective Influence and DomiRank on a network consisting of a corona of hubs.} A comparison between (a) Collective Influence and (b) DomiRank scores to highlight their differences in evaluating nodal importance. Note that DomiRank is computed and displayed for $\sigma=\frac{0.75}{-\lambda_N}$, but for this topology, a similar pattern of the relative ranking of nodes is obtained for most of the range of $\sigma$ values.}
    \label{fig:SM_DRCI_Corona}
\end{figure*}

Another illuminating network configuration to display the differences between CI and DomiRank is displayed in Fig. \ref{fig:SM_DRCI_Corona}, where a corona of hubs is presented with the (a) CI and (b) DomiRank centralities displayed by the node coloring. This figure shows how CI is able to identify the central node as the most influential. From DomiRank's perspective, that same node is jointly dominated by its neighbors - i.e., the hierarchical corona of hubs. From a network-dismantling perspective, a DomiRank-based attack would sequentially remove the hubs surrounding the central node (similarly to the strategy followed for the 2D-Lattice), which satisfies two objectives: dismantling the hubs as fundamental parts of the structure of the network and continuously decreasing the influence of the central node (top influencer from the CI perspective) by isolating it. On the other hand, a CI-based attack would first remove the central node and then proceed with the removal of the hubs, which remain important due to their high degree.
The pragmatic strategy of  DomiRank, wherein it attempts to remove important influencers whilst decreasing the importance of the top influencer, yields competitive, if not superior, attack strategies in dismantling networks despite of not being recomputed sequentially after each nodal removal. 

\begin{figure*}[!h]
    \centering
    \includegraphics[width = \linewidth]{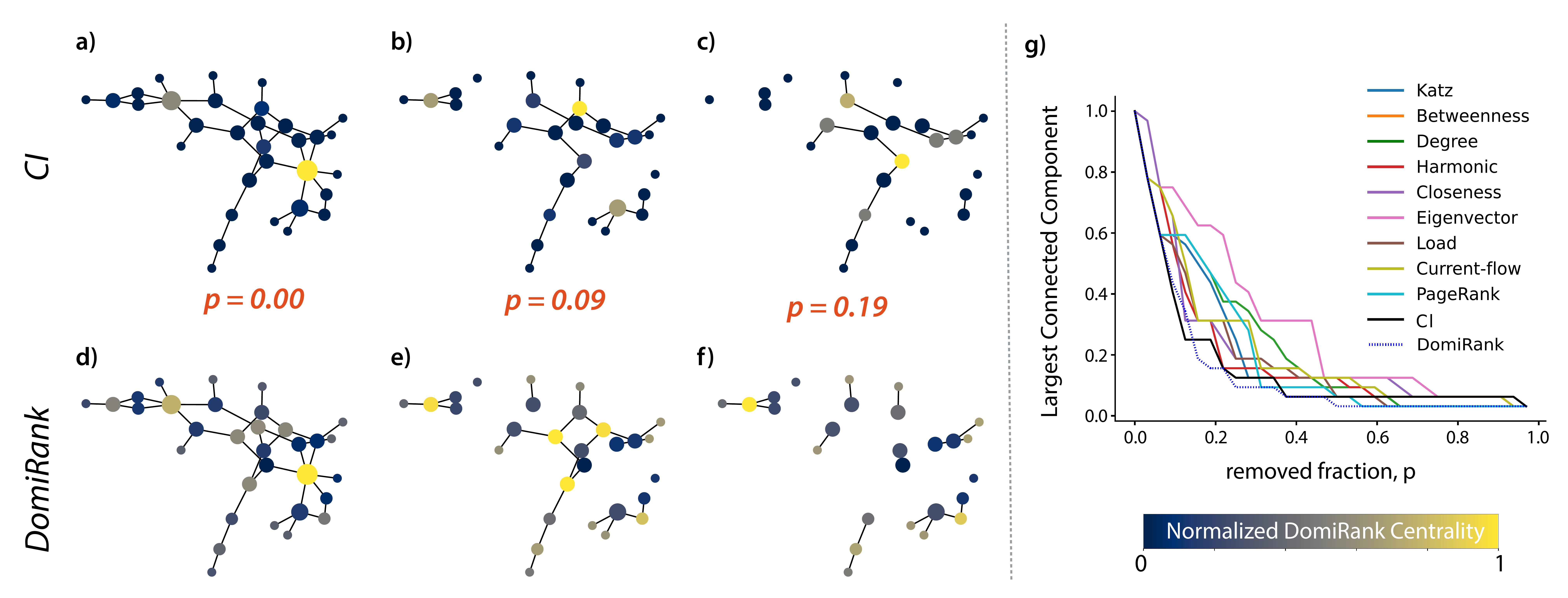}
    \caption{\textbf{Comparison of Collective Influence and DomiRank on a toy random network.} Comparison between (a,b,c) Collective Influence and (d,e,f) DomiRank and how they dismantle a toy Erd\H{o}s-R\'enyi network of size $N=32$ when nodes are sequentially removed according to descending values of their centralities. (g) The evolutions of the largest connected component of the network undergoing CI- and DomiRank-based attacks are displayed, along with various attacks based on other centralities for reference. Note that the color scale used to represent the relative values of the centralities in panels (a-f) is re-normalized for each panel  for visualization purposes }
    \label{fig:SM_DRCI_ToyER}
\end{figure*}

Yet another interesting example for the comparison of CI and DomiRank is the Erd\H{o}s-R\'enyi network in Fig. \ref{fig:SM_DRCI_ToyER}, where a mix of local and mesoscopic features might emerge. Here, Fig. \ref{fig:SM_DRCI_ToyER}(a,d) shows that CI and DomiRank agree on the identification of the two most important nodes despite their underlying mechanistic differences. However, the relative importance of nodes differs for the other intermediate nodes. Particularly,  Fig. \ref{fig:SM_DRCI_ToyER}(b,e) shows a similar phenomenon as highlighted by Figs. \ref{fig:SM_DRCI_ToyGRID}, \ref{fig:SM_DRCI_Corona}, that the most important node, according to CI, is one of the least important nodes for DomiRank, as DomiRank aims to isolate those nodes by removing their periphery. This effect is further shown in the temporal evolution displayed in Fig. \ref{fig:SM_DRCI_ToyER}(d-f), where DomiRank-based attacks tend to locally fragment the network's sub-structures in their attempt to dismantle the overall network.  Finally, Fig. \ref{fig:SM_DRCI_ToyER}g shows that both of these methodologies are efficient at dismantling this toy network, slightly outperforming one another at different time steps, whilst being substantially better than other metrics at generating attacks to dismantle the network.

\section{Link removal during attacks}
\begin{figure*}
    \centering
    \includegraphics[width =0.8\linewidth]{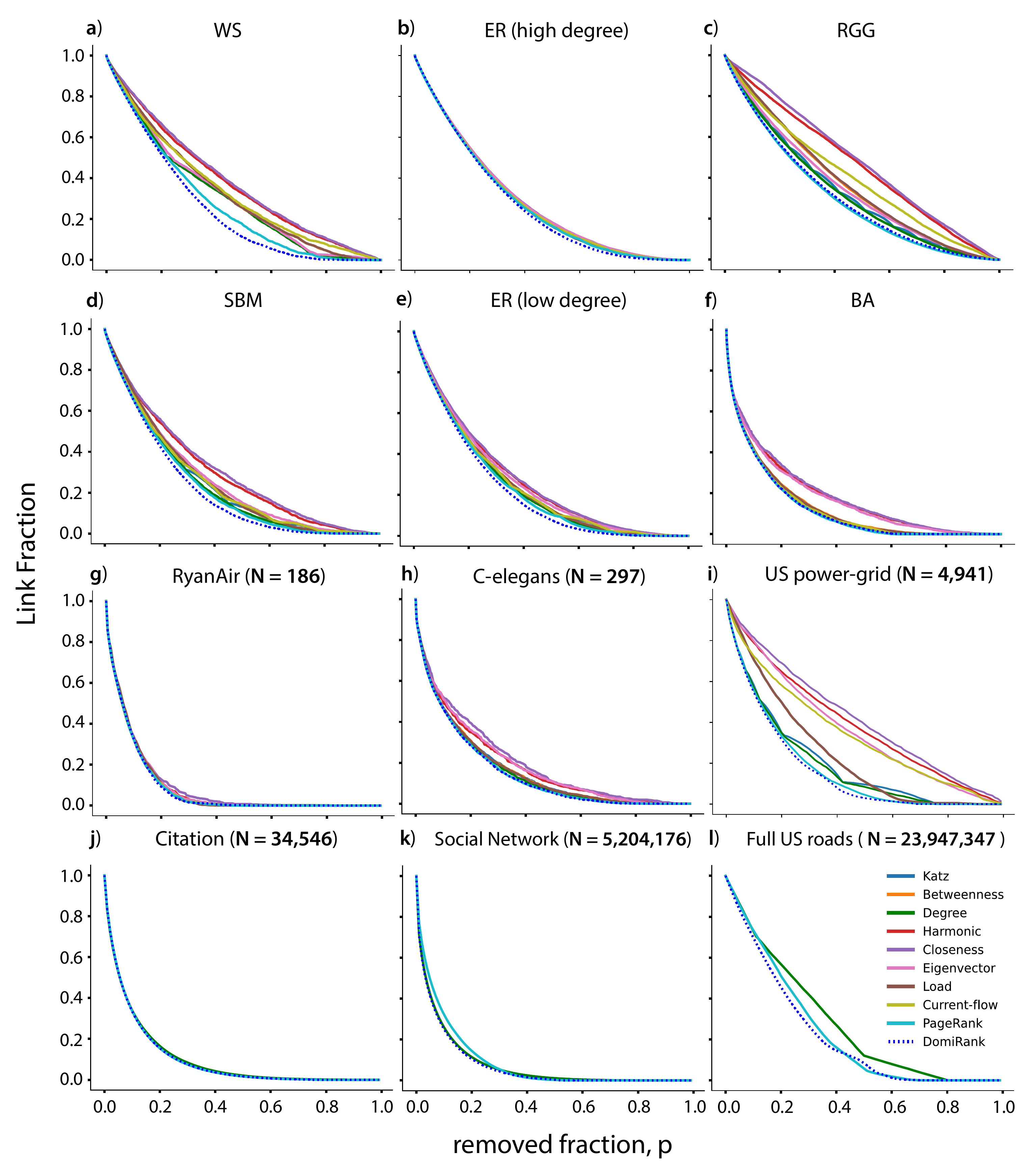}
    \caption{\textbf{Link-removal on synthetic and real-world networks under centrality-based attacks.} Evolution of the remaining link fraction whilst undergoing sequential node removal according to descending scores of various centrality measures for different synthetic networks of size $N=1000$: (a) Watts-Strogatz (WS; $\bar{k} = 4$), Erd\H{o}s-R\'enyi (ER) with (b) high ($\bar{k} = 20$) and (e) low link density ($\bar{k} = 6$), (c) random geometric graph (RGG; $\bar{k} = 16$), (d) stochastic block model (SBM; $\bar{k} = 7$), and (f) Bar{\'a}basi-Albert (BA; $\bar{k} = 6$). The performance of the attacks based on the different centrality metrics is also shown for different real networks: (g) hub-dominated transport network (airline connections, $\bar{k} = 16$), (h) neural network (worm, $\bar{k} = 29$), (i) spatial network (power-grid, $\bar{k} = 3$) , (j) citation network ($\bar{k} = 25$), (k) massive social network ($\bar{k} = 19$), and (l) massive spatial transport network (roads, $\bar{k} = 5$). Note that for panels j, k, and l, where massive networks are shown, only a few attack strategies are displayed due to the impossibility of computation of the rest.
    }
    \label{fig:SMLinks}
\end{figure*}

Our results have shown that attack strategies based on DomiRank centrality are more efficient in deteriorating the connectivity of the network in terms of the largest connected component than any other centrality-based attack. In this section (see Fig. \ref{fig:SMLinks}), we also show that DomiRank-based attacks are able to remove links more efficiently than other attacks for synthetic and real-world networks, providing further evidence about the capacity of the DomiRank to highlight the nodes structurally important for the integrity of the network's connectivity. 

The efficiency of DomiRank-based attacks in deteriorating network structural connectivity, both in terms of the largest connected component and sparsifying the number of connections, underlies the also outstanding capacity of this attack to severely impair the functionality of networks (see main text Fig. 6).

\section{Networks undergoing random-recovery}
\begin{figure*}
    \centering
    \includegraphics[width =\linewidth]{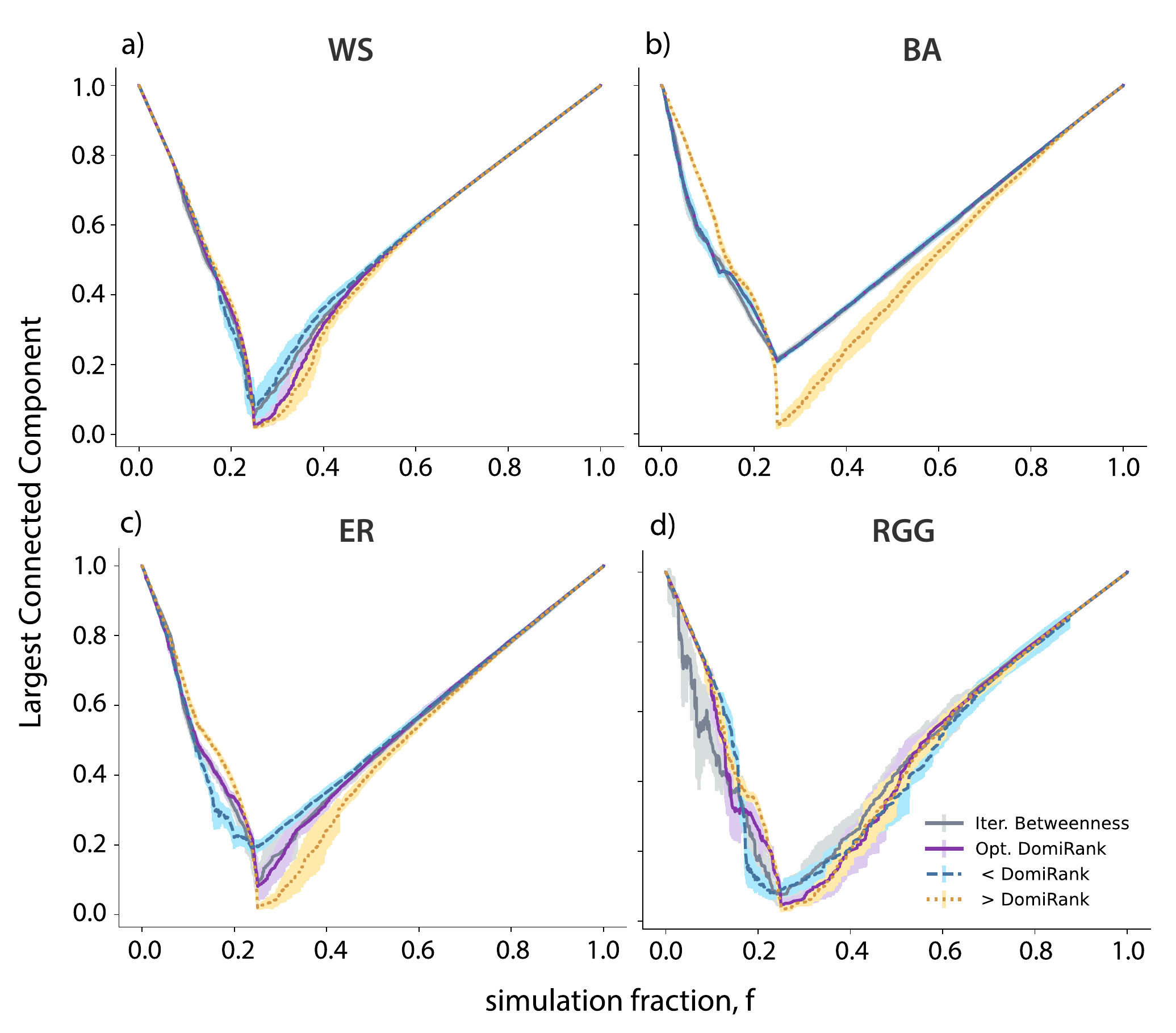}
    \caption{\textbf{Evaluating the effect of random recovery during iterative betweenness-based and DomiRank attacks.} Evolution of the relative size of the largest connected component for various synthetic networks of size $N=500$, namely (a) Watts-Strogatz (WS; $\bar{k}=4$), (b) Bar{\'a}basi-Albert (BA; $\bar{k}=6$), (c) Erd\H{o}s-R\'enyi (ER; $\bar{k}=5$), and (d) random geometric graph (RGG; $\bar{k}=7$), undergoing sequential node removal based on pre-computed DomiRank (optimal, low ($<$), and high ($>$) $\sigma$) and iterative betweenness, while a random node recovery process is ongoing; specifically, at each time step there is a probability of $0.25$ to recover a random (already removed) node.
    }
    \label{fig:SMRandom}
\end{figure*}

Complementing the results shown in Figure 5 in the main text, we show how a different recovery mechanism affects the evolution of the largest connected component under various attacks. Particularly we implement a random recovery mechanism, for which at every time step, a node is selected at random (with uniform probability) from the pool of the removed nodes. This selected node is recovered with probability $p$.  Note that if a given node is recovered, it cannot be subject to any further attack. Our results using a random recovery mechanism are consistent with those shown in Figure 5 in the main text, where a sequential recovery mechanism was implemented. Notably, the high-$\sigma$ DomiRank-based attack (Fig. \ref{fig:SMRandom}) inflicts more enduring damage (i.e., longer time to recover the same relative size of the largest connected component) than the iterative betweenness when the network has a random recovery process, despite the fact that the iterative betweenness-based-attacks are superior in dismantling the structure of the network. Fig. \ref{fig:SMRandom}a-d also showcases that when a network has a random recovery process, a low-$\sigma$ DomiRank-centrality-based attack can result in an incrementally more rapid deterioration of the largest-connected-component than iterative betweenness. This fundamentally shows a key property of DomiRank-based attacks, i.e., the inherent trade-off between the efficiency of network dismantling and the endurance of the damage by modulating the parameter $\sigma$ from low to high.

\newpage
%